\documentclass[aps,prd,twocolumn,superscriptaddress,nofootinbib]{revtex4-2}

\usepackage{graphicx}%
\usepackage{amsmath,amssymb,amsfonts}%
\usepackage{here}
\usepackage{xcolor}
\usepackage[normalem]{ulem}
\usepackage{hyperref}
\usepackage{float}%
\usepackage{braket}

\begin{document}

\title{Quantum vs. semiclassical description of in-QGP quarkonia in the quantum Brownian regime}

\author{Aoumeur Daddi Hammou}
\email[]{daddiham@subatech.in2p3.fr}
\affiliation{SUBATECH, IMT Atlantique, Nantes Université, CNRS/IN2P3,  Nantes, 44307, France}

\author{Stéphane Delorme}
\email[]{stephane.delorme@us.edu.pl}
\affiliation{Institute of Physics -- University of Silesia, Katowice, Poland}
\affiliation{Institute of Nuclear Physics, Polish Academy of Sciences, 31-342 Kraków, Poland}

\author{Jean-Paul Blaizot}
\email[]{jean-paul.blaizot@ipht.fr}
\affiliation{Institut de physique théorique, Université Paris-Saclay, CNRS, CEA, Gif-sur-Yvette 91191, France}

\author{Pol Bernard Gossiaux}
\email[]{gossiaux@subatech.in2p3.fr}
\affiliation{SUBATECH, IMT Atlantique, Nantes Université, CNRS/IN2P3,  Nantes, 44307, France}

\author{Thierry Gousset}
\email[]{gousset@subatech.in2p3.fr}
\affiliation{SUBATECH, IMT Atlantique, Nantes Université, CNRS/IN2P3,  Nantes, 44307, France}

\date{\today}

\begin{abstract}
In this work, we explore the range of validity of the semiclassical approximation of a quantum master equation designed to describe the $c\bar{c}$ dynamics in a quark-gluon plasma at various temperatures, in the quantum Brownian regime. We perform a comparative study of various properties, e.g. the charmonia yield,  of the Wigner density obtained both with the Lindblad equation and with the associated semiclassical Fokker-Planck equation. The semiclassical description is found to reproduce with a remarkable accuracy the results obtained through the full quantum description. We show that, to a large extent, this can be attributed to the non-unitary components of the dynamics that result from the contact of the $c\bar{c}$ subsystem with the thermal quark-gluon bath, leading to a rapid classicalization of the subsystem.
\end{abstract}

\keywords{Quark-Gluon Plasma, Quarkonia, Open Quantum Systems, Semiclassical Approximation}

\maketitle
\tableofcontents

\section{Introduction}

Quarkonia production in ultrarelativistic heavy ions collisions (URHIC) is considered as one of the standard probes of the quark-gluon plasma (QGP) created in such collisions for a short lapse of time~\cite{Wang:2016opj}. It has been  measured by several major experimental collaborations at the Relativistic Heavy Ion Collider (RHIC) and at the Large Hadron Collider (LHC) (see f.i. \cite{PHENIX:2006gsi,STAR:2022rpk,ALICE:2023hou,CMS:2016rpc,LHCb:2021bfl}). Such measurements reveal the impact of the quark-gluon plasma  on the short-distance ($\approx 1\,{\rm fm}$) interaction between two color charges. Indeed, the vast majority of effective theories and models designed to describe the quarkonia yield in URHIC are of a dynamical nature and rely both on the local QGP properties (such as its temperature) and on the effective interaction induced by the QGP  between an heavy quark $Q$ and its antiquark $\bar{Q}$ partner, either explicitly in microscopic approaches \cite{Young:2008he,Akamatsu:2012vt,blaizot2018quantum,Brambilla:2016wgg,Villar:2022sbv} or through the properties of in-medium quarkonia resonances, with their associated decay rates, in approaches based on hadronic degrees of freedom \cite{Du:2015wha, Liu:2009nb, Nendzig:2014qka}.\footnote{As an exception, the statistical approach, which has also been used in this context \cite{Andronic:2019wva}, only relies on the masses of various quarkonia states.} Lattice QCD calculations achieve regular progresses in providing valuable inputs and constrains for these models in terms of the spatial diffusion coefficient $\kappa$ \cite{Altenkort:2023eav} or the complex potential $W$ ruling the dynamics of $Q\bar{Q}$ pairs \cite{Larsen:2024wgw}. However, in order to reach a precise interpretation of the experimental data on quarkonia yields and flows, a suitable transport theory should be identified and faithfully implemented. 

While many treatments in the past have relied on {\it bona fide} semiclassical (SC) formulations like rate equations \cite{Du:2015wha, Liu:2009nb, Nendzig:2014qka} describing the formation and destruction of quarkonia states or Fokker–Planck equations \cite{Young:2008he} describing the evolution of heavy quarks in phase space, the open quantum system framework~\cite{breuer2002theory,rivas2012open,joos2013decoherence} has emerged in the heavy flavor community~\cite{Young:2010jq,Akamatsu:2012vt,Katz:2015qja,Blaizot:2015hya}  as a more rigorous first-principle approach allowing us to preserve some essential quantum features of the dynamics. Based on the hierarchy between various time scales pertaining to the QGP medium and the $Q\bar{Q}$ pair (see e.g. \cite{akamatsu2022quarkonium,Yao:2021lus} and references therein), several regimes can be identified: At “high” temperature ($T\gtrsim \Delta E$, where $\Delta E$ is the typical energy difference between bound states), the $Q\bar{Q}$ Heisenberg time $1/\Delta E$ is larger than the autocorrelation time of the QGP ($\propto 1/T$), which mostly acts as a white noise on the pair. This is the so-called “quantum Brownian regime” (QBM) for which the reduced density matrix follows a Quantum Master Equation (QME) of the Gorini–Kossakowski–Sudarshan–Lindblad type \cite{Lindblad:1976,10.1063/1.522979}. As demonstrated in \cite{blaizot2018quantum}, this Lindblad equation is best formulated in space – momentum coordinates and admits a complex potential describing both the unitary and the non-unitary parts of the dynamics. At “low” temperature  ($T\lesssim \Delta E$), the QGP can be viewed as a perturbation inducing transitions between the singlet and the octet sectors of the $Q\bar{Q}$ pair, leading to a QME for the density matrix formulated on the eigenbasis of the real part of the interaction and implying generalized transition rates between genuine weights (diagonal matrix elements  $\langle n |\mathcal{D}| n \rangle $) and “quantum coherence” (matrix elements of the  $\langle n |\mathcal{D}| m \ne n \rangle $ type) \cite{Yao:2018nmy}.      

To our knowledge, a full numerical implementation of the QME preserving the color structure has only been achieved in the QBM and for a single $Q\bar{Q}$ pair \cite{Miura:2022arv, Brambilla:2023hkw,Delorme:2024rdo}. This allowed direct phenomenological applications for Upsilon yield, the initial creation of a unique $b\bar{b}$ pair being the most frequent occurrence, even for central Pb-Pb collisions at LHC energies. Adopting the alternative quantum optical regime (QOR) perspective to describe Upsilon production operationally \cite{Yao:2020xzw} could only be achieved by first resorting to the SC approximation which consists, for such regime, in neglecting the coherences in the QME, resulting in usual rate equations. Although identified formally in \cite{Yao:2020eqy}, quantum corrections await for concrete evaluation.

When facing the charmonia production in AA collisions, given that up to 100 $c\bar{c}$ pairs are generated during the initial stages, a full quantum treatment seems at present out of reach. In this context, the use of  semiclassical equations is  attractive due to their simplicity and numerical effectiveness as compared to the QME. They enable us to  study the simultaneous evolution of several $c\bar{c}$ pairs and get insights on the recombination  aspect of  the in-QGP quarkonia dynamics. For the QBM regime ruling the first stage of the QGP evolution, it was shown in \cite{blaizot2018quantum,akamatsu2022quarkonium} that these SC equations take the form of coupled Klein-Kramers equations, each one describing the evolution of a given color representation in the heavy sector, while a first application to the case of several $c\bar{c}$ pairs was discussed therein. Since this seminal work, other calculations following the same spirit have been performed in more realistic conditions \cite{Villar:2022sbv}, pursuing the series of models dealing with the study of in-QGP quarkonia (see \cite{andronic:2024oxz} and references cited therein). 

For closed systems, the validity of the semiclassical approximation  hinges on certain generic conditions stating that the wavelength of the particles should be much smaller than the scale over which the potential varies and that the action $S$ should be large compared to $\hbar$. A direct numerical application reveals that these conditions are not implicitly satisfied when evolving and/or describing the lowest bound states of a $Q\bar{Q}$ system, for which $S\sim \hbar$. For quantum (sub)systems coupled with some environment at finite temperature, it is known that the environment will gradually quench the quantum properties of the subsystem, leading to a classicalization process associated to the transition from a pure initial state to a mixed state (see f.i. \cite{heller1976wigner, schlosshauer2007decoherence, joos2013decoherence} and references therein), thus legitimating the use of semiclassical equations in a second stage of the evolution. For practical purposes the essential questions are therefore to assess the typical time after which a system composed of heavy quarks and antiquarks can be described with semiclassical equations as well as the numerical uncertainty resulting from this treatment. In order to explore the range of validity of the semiclassical approximation and attempt to answer these questions, we conduct in this work a comparative analysis between the results obtained from a quantum master equation  and its approximate semiclassical equation, focusing on a single $c\bar{c}$ pair in the QBM regime. In fact, as discussed in \cite{blaizot2018quantum}, this approximation is best implemented in the abelian or quantum electrodynamics (QED) version of the QME, as color to singlet transitions imply finite energy exchanges and are genuinely of quantum nature. We thus restrict ourselves to this abelian case and plan to address the QCD case in a future work. 

In section \ref{sect:background}, we provide a brief reminder of the theoretical background. The general strategy for the comparison is presented in section \ref{Generalstrategy}. We next present our essential results, first addressing the global picture of the $c\bar{c}$ evolution in section \ref{section_global} and then more specific quarkonia observables in section \ref{local-comparaison}. Section \ref{conclusions} contains our conclusions. 

\section{Theoretical background}
\label{sect:background}

Hereafter, we summarize the essential ingredients of recent works dedicated to Quantum Master Equations (QME) in the quantum Brownian regime \cite{blaizot2018quantum,Delorme:2024rdo}
as well as their semiclassical approximation. The equations derived in \cite{blaizot2018quantum} exploit the non relativistic character of the heavy quark ($Q$) dynamics, and are supposed to be valid in a regime where the temperature, while remaining small compared to the heavy quark mass ($T\ll M$) is large compared to the quarkonium binding energy $E$, or the energy differences $\Delta E$ between  quarkonia states, i.e., $T\gg E\gtrsim \Delta E$. The resulting master equations  are conveniently written as  
\begin{equation}
	\frac{\text{d}{\mathcal{D}}\left(t\right)}{\text{dt}}={\mathcal{L}}\left[{\mathcal{D}}\left(t\right)\right]=\sum_{i=0}^4 {\mathcal{L}}_i{\mathcal{D}},\label{Lindblad-map}
\end{equation}
where ${\mathcal{D}}$ is the reduced density matrix of the $Q\bar{Q}$ pair, and  ${\mathcal{L}}$ is the total Liouville superoperator  that generates evolution  in time.   The operator ${\cal L}$ can be decomposed,  into a set of five  superoperators ${\mathcal{L}}_i$  which  capture different aspects of the dynamics (to leading orders in the time derivatives). Their generic expressions are given by 
\begin{equation}
	\begin{array}{cclc}
		{\mathcal{L}}_{0}{\mathcal{D}} & \equiv & -i\text{\ensuremath{\left[{H},{\mathcal{D}}\right]},}\\
		\\
		{\mathcal{L}}_{1}{\mathcal{D}} & \equiv & -\frac{i}{2}\int_{  {xx^{\prime}}}V\left(  {x-x^{\prime}}\right)\text{\ensuremath{\left[{n}_{  {x}}{n}_{  {x^{\prime}}},{\mathcal{D}}\right]},}\\
		\\
		{\mathcal{L}}_{2}{\mathcal{D}} & \equiv & \frac{1}{2}\int_{  {xx^{\prime}}}W\left(  {x-x^{\prime}}\right)\text{\ensuremath{\left(\left\{  {n}_{  {x}}{n}_{  {x^{\prime}}},{\mathcal{D}}\right\}  -2{n}_{  {x}}{\mathcal{D}}{n}_{  {x^{\prime}}}\right)},}\\
		\\
		\mathcal{{L}}_{3}\mathcal{D} & \equiv & -\frac{i}{8T}\int_{xx'} W\left(x-x'\right)
		\left(
		\left\{  \mathcal{D},\left[\dot{n}_{x'},n_x\right]\right\}  +2\dot{n}_{x'}\mathcal{D} n_x 
		\right.\\	
		&&\left.-2{n}_{x}{\mathcal{D}}\dot{n}_{  {x^{\prime}}}\right),\\
		\\
		{\mathcal{L}}_{4}{\mathcal{D}} & \equiv & \frac{1}{32T^{2}}\int_{  {xx^{\prime}}}W\left(  {x-x^{\prime}}\right)\text{\ensuremath{\left(\left\{  \dot{{n}}_{  {x}}\dot{{n}}_{  {x^{\prime}}},{\mathcal{D}}\right\}  -2\dot{{n}}_{  {x}}{\mathcal{D}}\dot{{n}}_{  {x^{\prime}}}\right)},}
	\end{array}\label{eq:equatQME}
\end{equation}
where ${H}$ is  the kinetic energy  of the heavy quarks, $V$ and $W$ are respectively the real and imaginary parts of a complex potential,  and ${n}_{  {x}}$ is the charge density operator, with $\dot n_x$ its time derivative. Both $V$ and $W$ depend on the medium through which the heavy quarks are propagating. The real potential $V$ includes the screening corrections, while $W$ represents generically the effect of collisions of the heavy quarks with the medium constituents. The operators  ${\mathcal{L}}_{0}$ and ${\mathcal{L}}_{1}$ produce a unitary (or coherent) evolution, while the other superoperators ${\mathcal{L}}_{2,3,4}$, proportional to  $W$,  are responsible for the non-unitary dynamics, with the associated phenomena of decoherence and dissipation. The superoperator $\mathcal{L}_4$ ensures that Eq.~(\ref{Lindblad-map}) takes the form of a Lindblad equation, thereby guaranteeing that the evolution preserves the strict positivity of the density matrix. As discussed at length in \cite{Delorme:2024rdo} it is however subleading with respect to the other operators (it contributes for instance a correction to the diffusion constant which is quadratic in the heavy quark velocity). Furthermore, in the semi-classical approximation it plays no role in the positivity.

The solutions of the equations (\ref{eq:equatQME}), for the case a single   $Q\bar{Q}$ pair in a one-dimensional setting, were obtained  and  analyzed in  \cite{Delorme:2024rdo}, where explicit expressions of the various superoperators can be found as well. In the present analysis, we restrict ourselves to the abelian case, and furthermore we focus on the relative motion. The explicit expressions of the superoperators obtained  after tracing out the center of mass coordinates are given in Appendix A. 

We denote by $s$  the relative coordinates of the quarks, and by ${\cal D}(s,s')$ the matrix element $\bra{s}{\cal D}\ket{s'}$. For implementing the semiclassical approximation, it is convenient to introduce the following coordinates $r=(s+s')/2$ and $y=s-s'$.  The Fourier transform of ${\cal D}$ with respect to $y$ (for fixed $r$)
\begin{eqnarray}\label{eq:Wignerdef}
	\mathcal{D}\left( r, p\right)=
	\int\text{d} y\,  e^{-\frac{i p  y}{\hbar }}\mathcal{D}(r+\frac{  y}{2},r-\frac{  y}{2}), 
\end{eqnarray}
is the Wigner transform. It is denoted,  with a slight abuse of notation, by  ${\cal D}(r,p)$.

As was shown in \cite{blaizot2018quantum}, in the semiclassical approximation, which relies in particular on the fact that $\mathcal{D}(r,y)$ is peaked around $y=0$,  Eq.~(\ref{Lindblad-map}) yields  the following  Fokker-Planck equation: 
\begin{equation}
	\frac{\partial \mathcal{D}}{\partial t} =  \left[-\frac{2{p}{\partial}_{{r}}}{M}-V'(r)\partial_p+\frac{\eta\left({r}\right)}{2}\partial_p^2+\frac{2\gamma\left({r}\right)}{M}\partial_{p} p\right]\mathcal{D},
	\label{Fokker}
\end{equation}
where 
\begin{equation}
	\eta(r)=\frac{C_F}{2}(\tilde{W}''(0)+\tilde{W}''(r))
	\quad\text{and}\quad \gamma(r)=\frac{\eta(r)}{2T},
	\label{eq:defcoeff}
\end{equation}
satisfying the Einstein relation. It is worth to note that in deriving Eq. (\ref{Fokker})  from the Lindblad equation (\ref{Lindblad-map}), the operator $\mathcal{L}_4$ was not included. As mentioned above, ${\mathcal{L}}_4$ is intrinsically subdominant and unnecessary to warrant positivity in the SC approximation. 

It can be shown that the steady state solution of Eq.~(\ref{Fokker}) is a Gibbs-Boltzmann distribution.  This is also the case for the quantum dynamics as  described by Eq.~(\ref{Lindblad-map}), but with an effective temperature differing slightly from the real medium temperature.\footnote{See \cite{Delorme:2024rdo} for a discussion of the equilibrium properties associated to the QME.}

\section{General strategy for the comparison} 
\label{Generalstrategy}

Our main goal in this paper is to provide a systematic comparison of the solutions obtained by solving the quantum evolution equations (\ref{Lindblad-map}) with those obtained by solving the Fokker-Planck equation (\ref{Fokker}), thereby getting quantitative measures of the validity of the semiclassical approximation (in the restricted context specified earlier: one-dimensional setting and abelian case). 
As the operator $\hat{\mathcal{L}}_{4}$ (see Eq.~(\ref{eq:equatQME})) was not included in the derivation of Eq.~(\ref{Fokker}) from Eq.~(\ref{Lindblad-map}),
the quantum results will be presented both with and without $\hat{\mathcal{L}}_{4}$ in the comparative study below, so defining an ``acceptability band" for the SC approximation.

In our comparative study  between the quantum and semiclassical descriptions, we follow the evolution of the Wigner transforms of a  ${\rm c}\bar{\rm c}$ pair coupled to a quark-gluon plasma at a given temperature.  In section~\ref{section_global}, we discuss general features of  this evolution, while in section~\ref{local-comparaison}  we focus on observables and regions of phase space that are most relevant to the physics of quarkonia. 

\paragraph{Setup of the comparison}
We  consider the evolution of a single $c\bar{c}$ state pair in  an abelian plasma, with $M$ chosen to be $1.469\,\mathrm{GeV}$.  The system is confined to 
a one dimensional box of  length 20~fm. The Lindblad QME (\ref{Lindblad-map}) and the corresponding Fokker-Planck equation  (\ref{Fokker}) are solved using in both approaches the same  real and imaginary potentials, respectively $V\left(r\right)$ and $W\left(r\right)$, that were derived and studied in detail in~\cite{Katz:2022fpb}  (see however Appendix B). These potentials were adjusted to reproduce  binding energies and damping rates that are in agreement with those deduced from some three-dimensional potential inspired by lattice data \cite{burnier2015quarkonium,lafferty2020improved,Katz:2022fpb}.

The evolution of the $c\bar{c}$ pair, and the quality of the semiclassical approximation, depend on the initial condition, on the time the pair spends in contact with the heat bath, on the heat bath temperature,  and also on the precise observable that is studied. The heat bath temperature plays a major role in driving the $c\bar{c}$ pair toward a semiclassical regime, via processes of decoherence and thermalization. To test the role of temperature, we present solutions of both QME and FP equations in a stationary QGP, using four different medium temperatures $T\in\left\{0.2,0.3,0.4,0.6\right\}$ in~GeV, the lowest of these temperatures being at the borderline between the quantum Brownian motion and the quantum optical regimes. We evolve the system up to a final time $t_{f}=20$~fm/$c$ where equilibration is nearly reached for all temperatures above $T=0.2\,\mathrm{GeV}$. 

For the initial state, we consider two options. As a first choice, we  take a Gaussian wave function 
\begin{equation}
	\psi\left(r\right)=\left(\frac{1}{\pi\sigma^{2}}\right)^{\frac{1}{4}}e^{-\frac{r^{2}}{2\sigma^{2}}} \label{eq:initial-state-def}  ,
\end{equation}
with a variance $\sigma=0.38$~fm optimized to reproduce at best the exact wave function of the $c \bar c$ lowest bound state in vacuum. In heavy ions collisions, quarkonia are ultimately produced from compact states in a position space basis.  Our second choice will therefore be that of an initial compact state with $\sigma=0.165$~fm.

\paragraph{The Lindblad equation as a benchmark} 
As discussed in \cite{Delorme:2024rdo}, the higher derivatives of the imaginary potential may lead to singular behaviors of the operators $\mathcal{L}_3$ and $\mathcal{L}_4$, unless the potential is properly regularized. We  use here the so-called ``minimal set'' of operators defined in \cite{Delorme:2024rdo} when solving the QME. This involves a modified imaginary potential  $\tilde{W}$, which, for consistency,  is also used to calculate the transport coefficients entering the FP equation.

The Lindblad QME (\ref{Lindblad-map}) is  solved  on a spatial 1D grid with  the numerical scheme established in \cite{delorme2021theoretical}, where details can be found.
The parameters used in the resolution are $\Delta r=0.04$ fm for the spatial step and $\Delta t=0.1$ fm/c for the time step. The Wigner representation (\ref{eq:Wignerdef}) of the density matrix is then obtained from the solution of the Lindblad equation  by taking a Fourier transform.

\paragraph{The Fokker Planck Langevin Equations}
The semiclassical expansion for the unitary operators $\mathcal{L}_0$ and $\mathcal{L}_1$  requires that the potential $V(r)$ be differentiable, which is not the case at the origin for the 1D potential derived in ~\cite{Katz:2022fpb}. We have therefore used, in the Langevin equation, Eq.~(\ref{eq:395-1}) below, a regularized potential $V_\mathrm{reg}$ which has a smooth behavior at the origin. This potential is presented in the Appendix B. Let us emphasize that this regularization, while essential for the SC calculation, has little impact on the quantum one.

With these ingredients, the Fokker-Planck equation (\ref{Fokker}) is solved numerically resorting to its equivalence with the  Langevin equations~\cite{blaizot2018quantum}, adopting a finite time step discretization: 
\begin{eqnarray}
	{ \rm d} r  &=&   \frac{p}{M_{\rm red}}\,{\rm d}t,\\
	{\rm d}p & = &   \left(- V'(r)-  \frac{\gamma(r)}{M_{\rm red}} p+  {\xi}\left(r\right) \right) {\rm d}t
	\label{eq:395-1},
\end{eqnarray}
where $M_{\rm red}=M/2$. The first equation represents the deterministic
motion of the relative position, while the second is a stochastic equation with the noise given by 
\begin{equation}
	\begin{array}{crl}
		\left\langle   {\xi}\left(r\right)\right\rangle = 0,\qquad 
		\left\langle   {\xi}\left(r\right)  {\xi}\left(r'\right)\right\rangle  = \delta(r-r') \frac{\eta(r)}{{\rm d}t},
	\end{array}
\end{equation}
with the drift and drag coefficients $\eta$ and $\gamma$ defined in Eq.~(\ref{eq:defcoeff}).
In practice, a time step ${\rm d} t=10^{-3}$ fm/c was used.\footnote{In order to ensure a better accuracy  for solutions, this time step gets divided by a factor of five for the small relative distances, namely,  $|r|\leq 0.2$ fm.  This reduction of the time step around the origin has been observed to reduce the numerical errors.} The expectation values of the various observables (including the Wigner functions) were obtained by averaging over an ensemble of $2\times 10^7$ simulations. The test-particle method was used to build the various Wigner distributions.

\section{Global picture of the $c\bar{c}$ evolution}
\label{section_global}
\begin{widetext}
	
	\begin{figure}[H]
		\centering
		\includegraphics[width=0.6\textwidth]{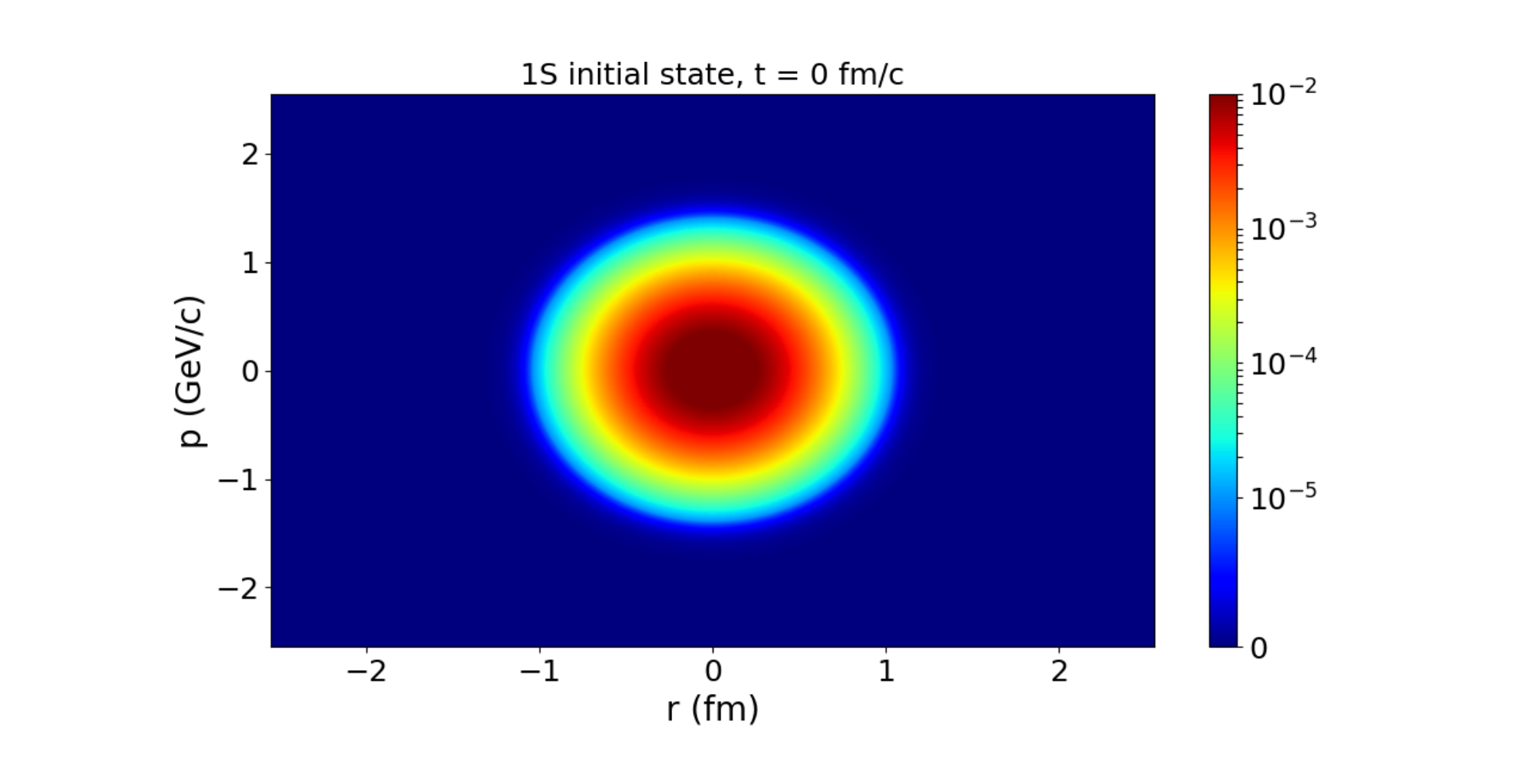}\\
		\includegraphics[width=0.32\linewidth]{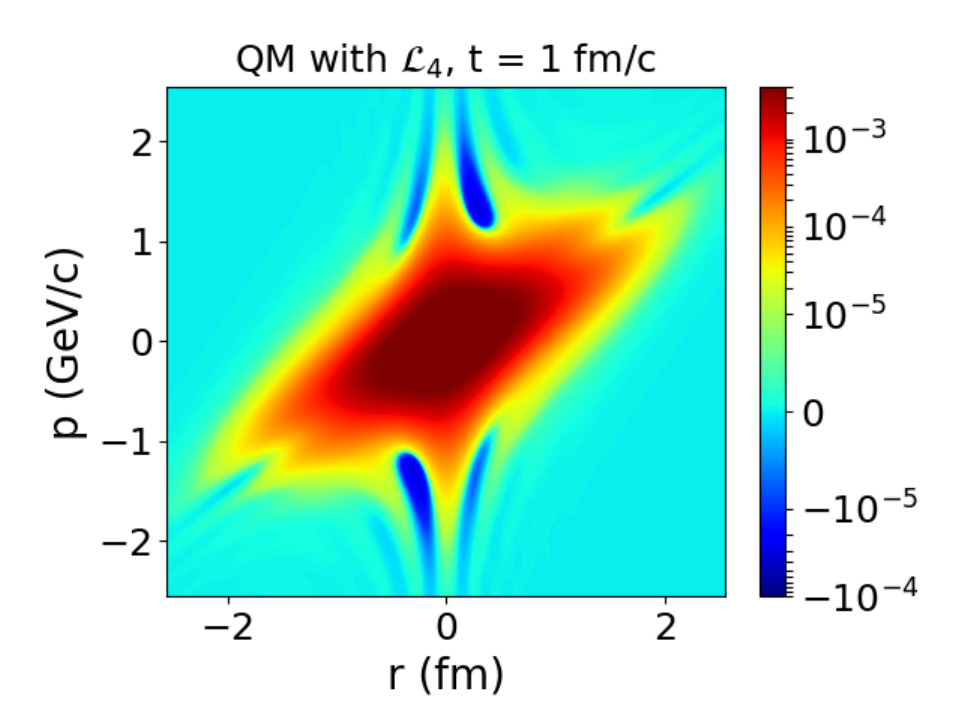}
		\includegraphics[width=0.32\linewidth]{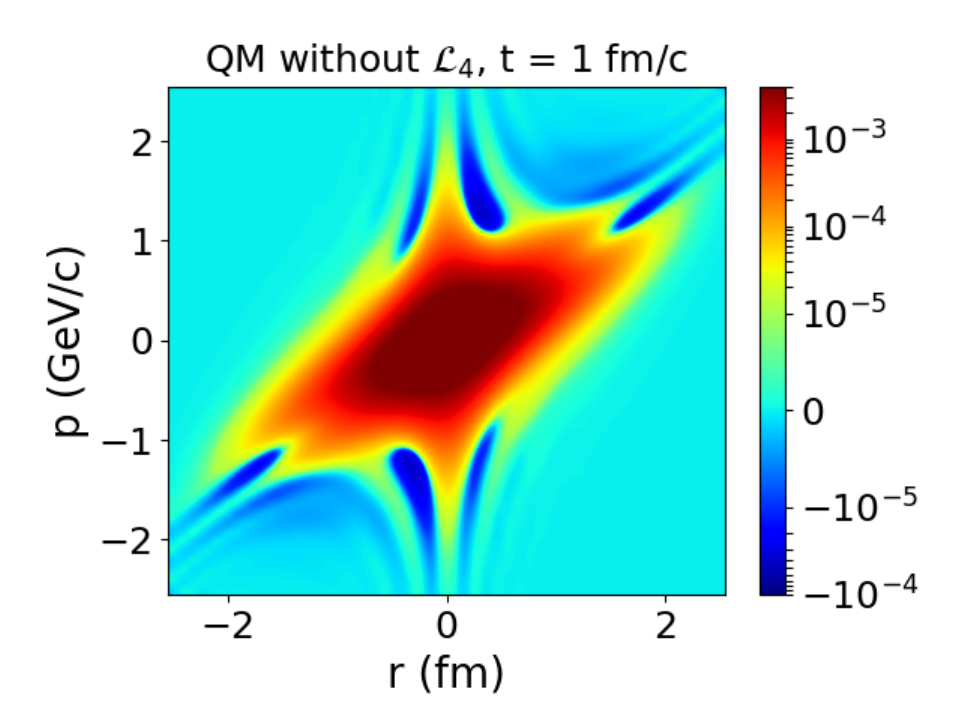}	
		\includegraphics[width=0.32\linewidth]{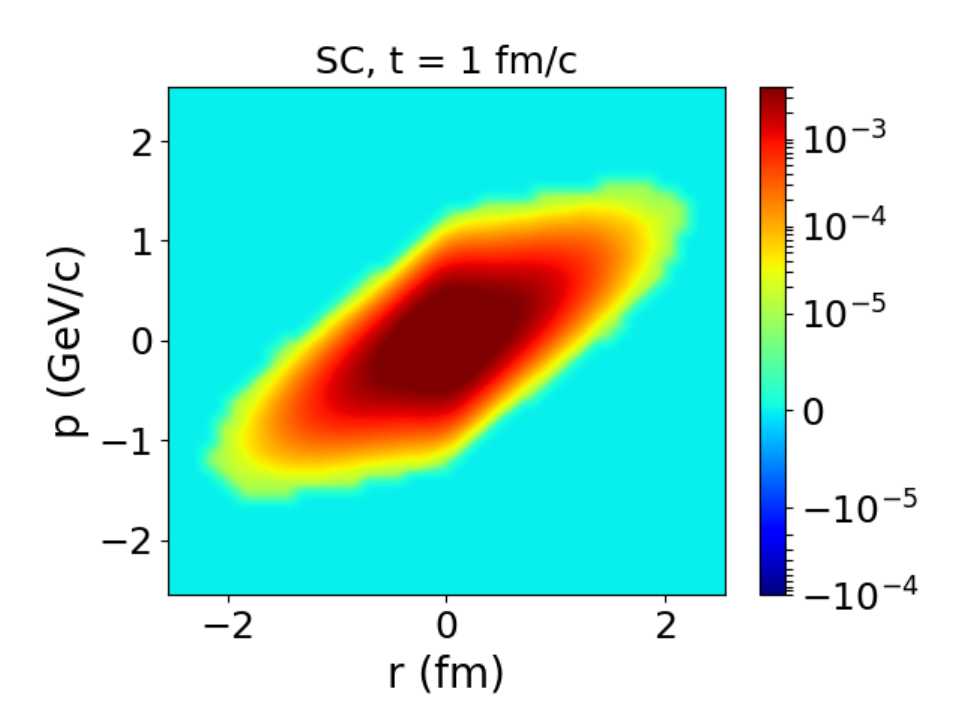}
		\includegraphics[width=0.32\linewidth]{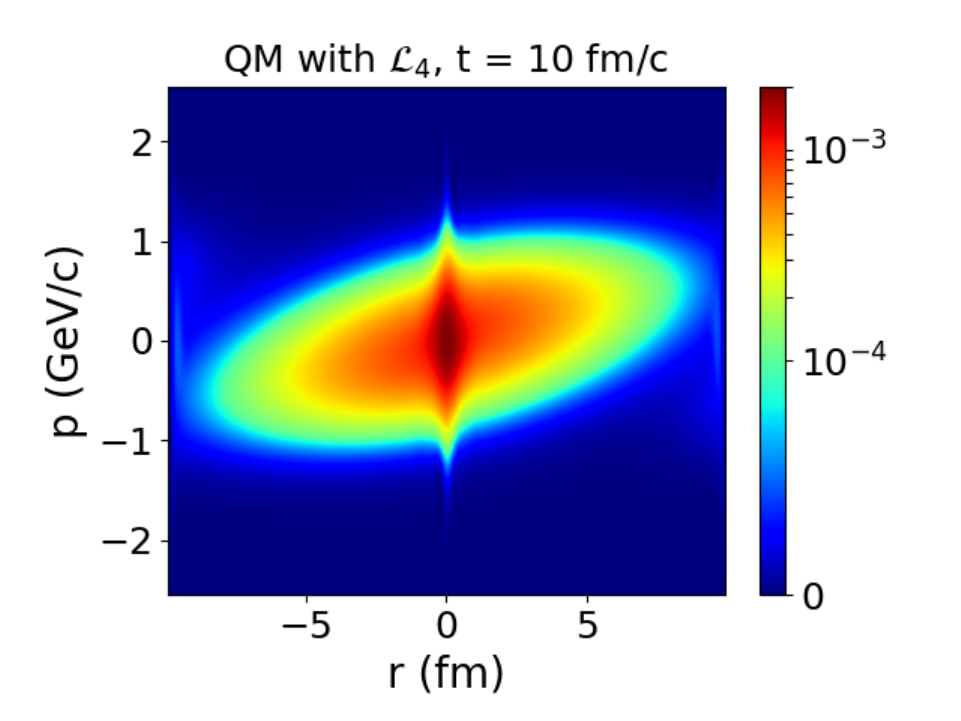}
		\includegraphics[width=0.32\linewidth]{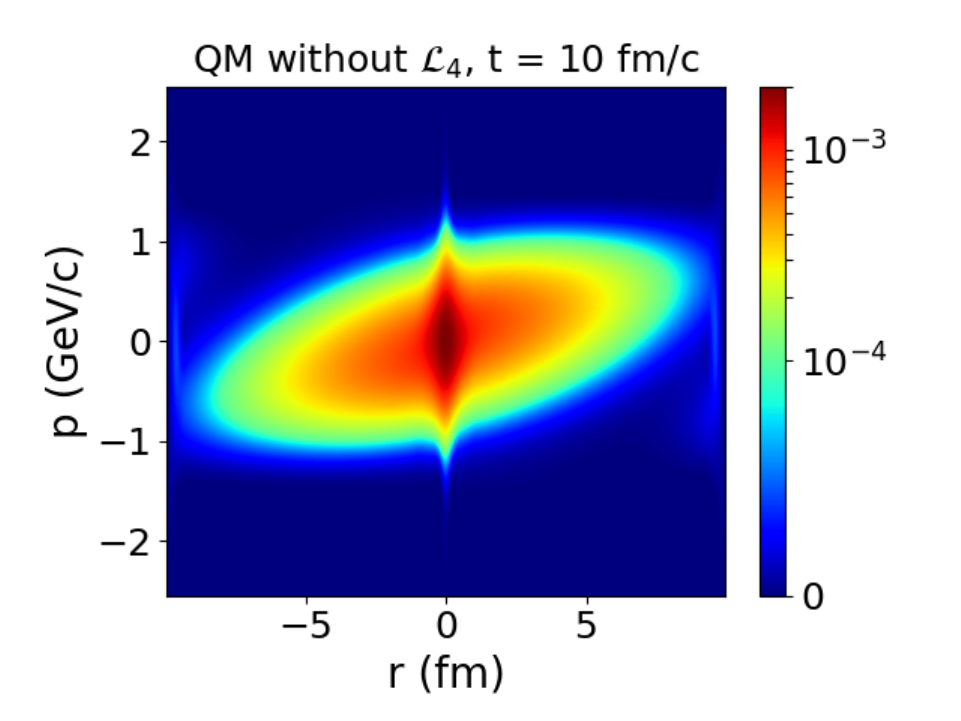}
		\includegraphics[width=0.32\linewidth]{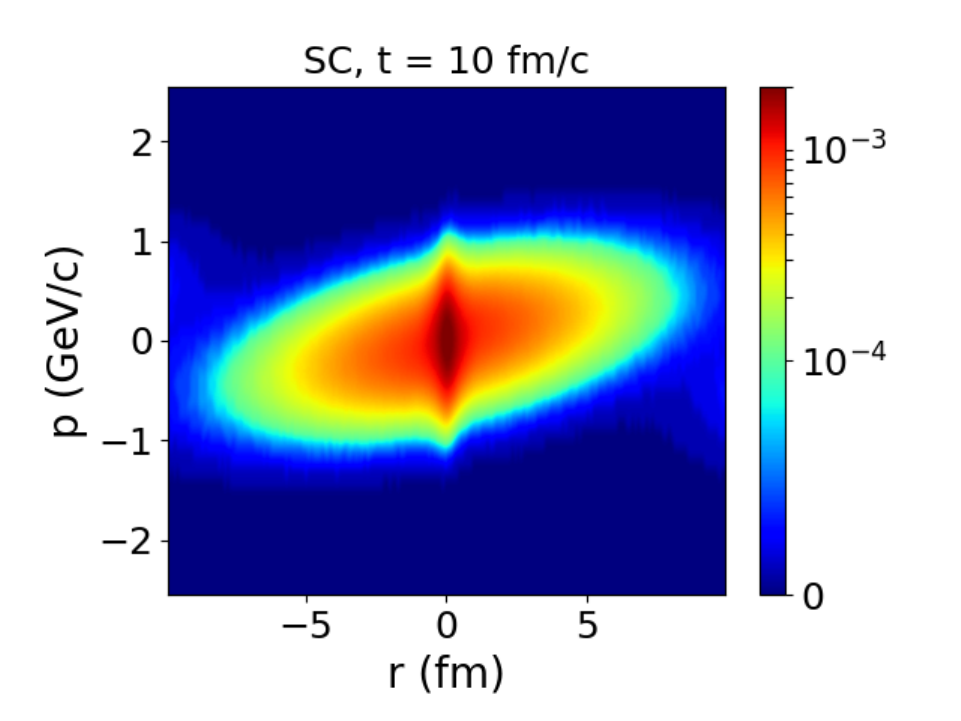}
		\includegraphics[width=0.32\linewidth]{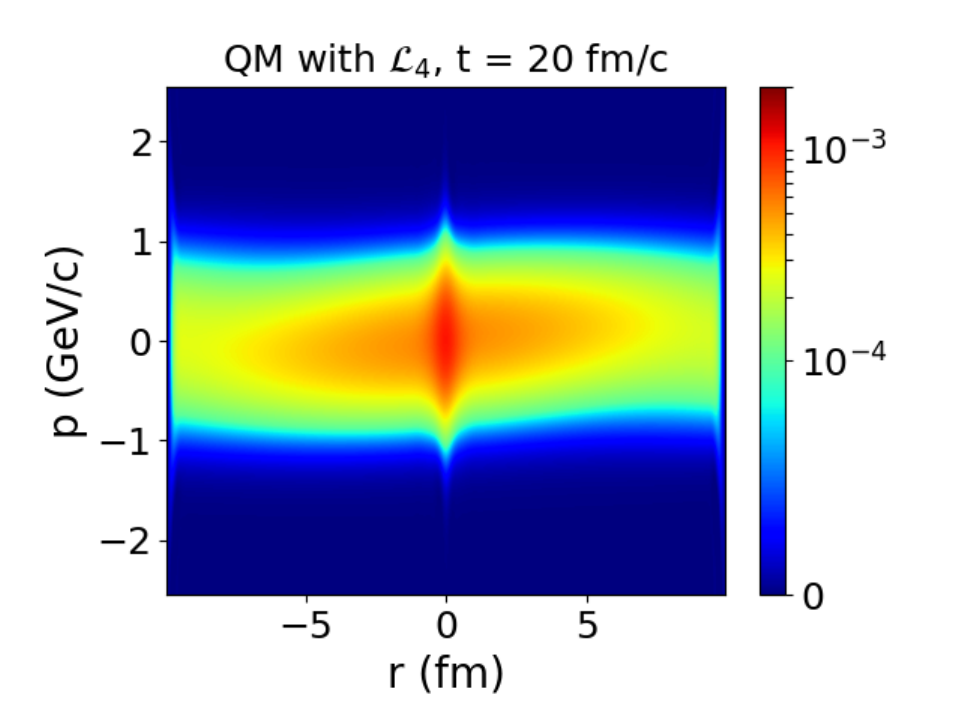}
		\includegraphics[width=0.32\linewidth]{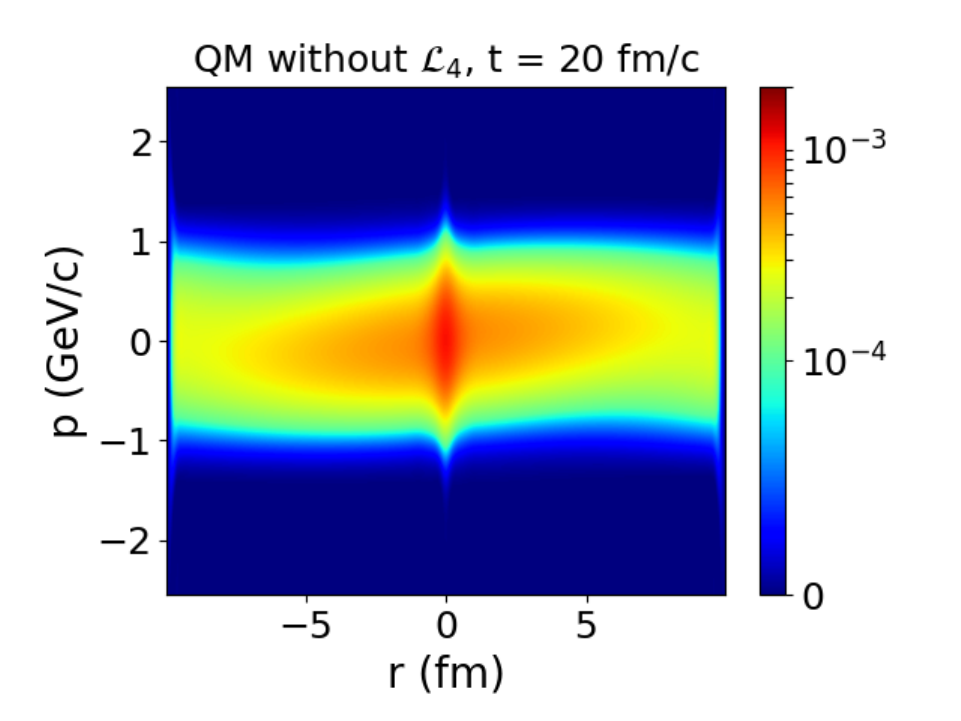}
		\includegraphics[width=0.32\linewidth]{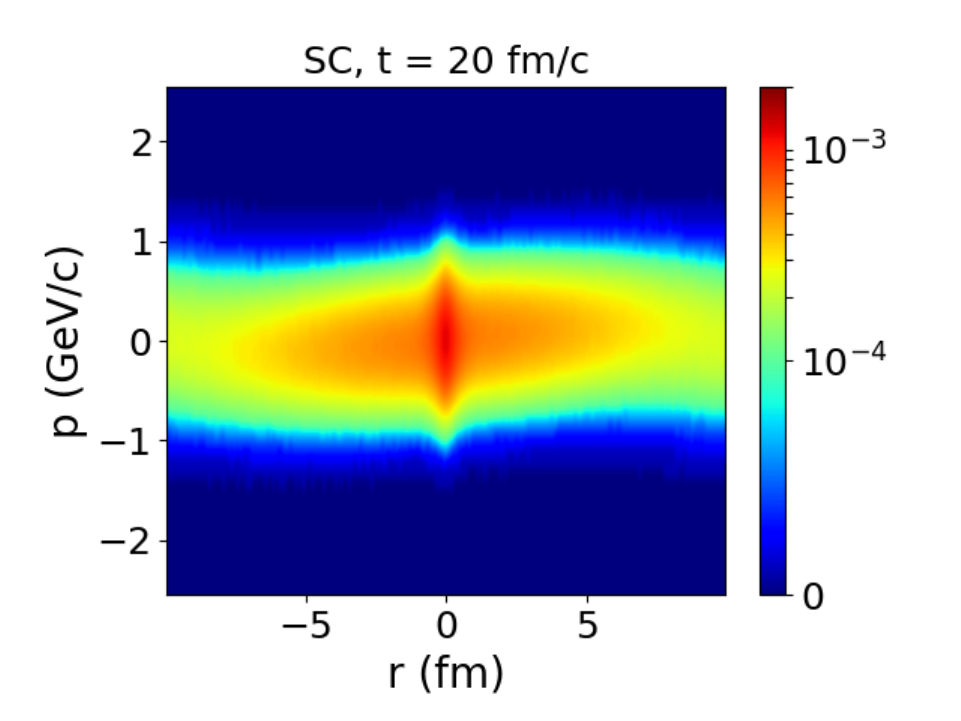}
		
		\caption{QM vs SC time evolution of the Wigner distribution $\mathcal{D}(r,p)$ for relative distance $r$ and momentum $p$ resulting from the evolution of a 1S initial state in a QGP at $T=0.3$~GeV. For the quantum case, we show the evolution both with (left column) or without (middle column) the $\mathcal{L}_4$ term.}\label{fig:snapshots-wigner}
	\end{figure} 
	
\end{widetext}

We start with an overall qualitative comparison of the Wigner transforms (\ref{eq:Wignerdef}) of the reduced density matrix of the ${c\bar{c}}$ pair, which is evolved in time either with the QME (\ref{Lindblad-map}) or by the FP equation (\ref{Fokker}). This is illustrated in Fig.~\ref{fig:snapshots-wigner}. The initial state is common to both evolutions: the Wigner transform of the Gaussian (\ref{eq:initial-state-def}), which is positive, and to which we refer as a ``1S" state with a slight abuse of notation.\footnote{While the first even excited state will be referred to as a ``2S" state.}  At early times, namely at $t=1$ fm/c, the dynamics are mainly unitary, driven by the operators $\mathcal{L}_0$ and $\mathcal{L}_1$. Because the 1S state is not an eigenstate of the in-medium potential, a non trivial evolution takes place: as the 1S state has a size smaller than that of the ground state of the in-medium potential, it starts expanding. This expansion is accompanied by quantum interferences which are clearly visible in the complex patterns in the Wigner distribution obtained from the quantum evolution. In contrast, the semiclassical evolution produces a smoother picture, but it reproduces faithfully the overall shape of the distribution reflecting the initial expansion of the 1S state. 
As time passes, the non-unitary component of the dynamics (induced by the operators ${\cal L}_2, {\cal L}_3,{\cal L}_4$) takes over and ensures an excellent agreement between the two descriptions toward the end of the evolution.

A comparison between the first two columns of Fig.~\ref{fig:snapshots-wigner} indicates the the operator $\mathcal{L}_4$ plays a minor role, except perhaps at the early stage of the evolution where it affects slightly the interference pattern.
Quite visible in Fig.~\ref{fig:snapshots-wigner} is also the survival of a bound state till late times, which manifests itself by the presence of a peak in the Wigner function near the origin of phase space. The correlation between $r$ and $p$, which reflects the initial motion of the heavy quarks, quickly disappears, and the momentum distribution acquires rapidly a symmetric shape at small $r$, slightly distorted at large $r$. At the same time, the density of unbound quarks extends so as to occupy the whole volume of the box, which is not achieved however until the last stages of the evolution. One then reaches a stationary regime, which looks remarkably similar in both the quantum and semiclassical descriptions. There are however quantitative differences that we shall now analyze more closely. 

\subsection{Thermalization}

\begin{figure}
	\centering
	\includegraphics[width=0.5\textwidth]{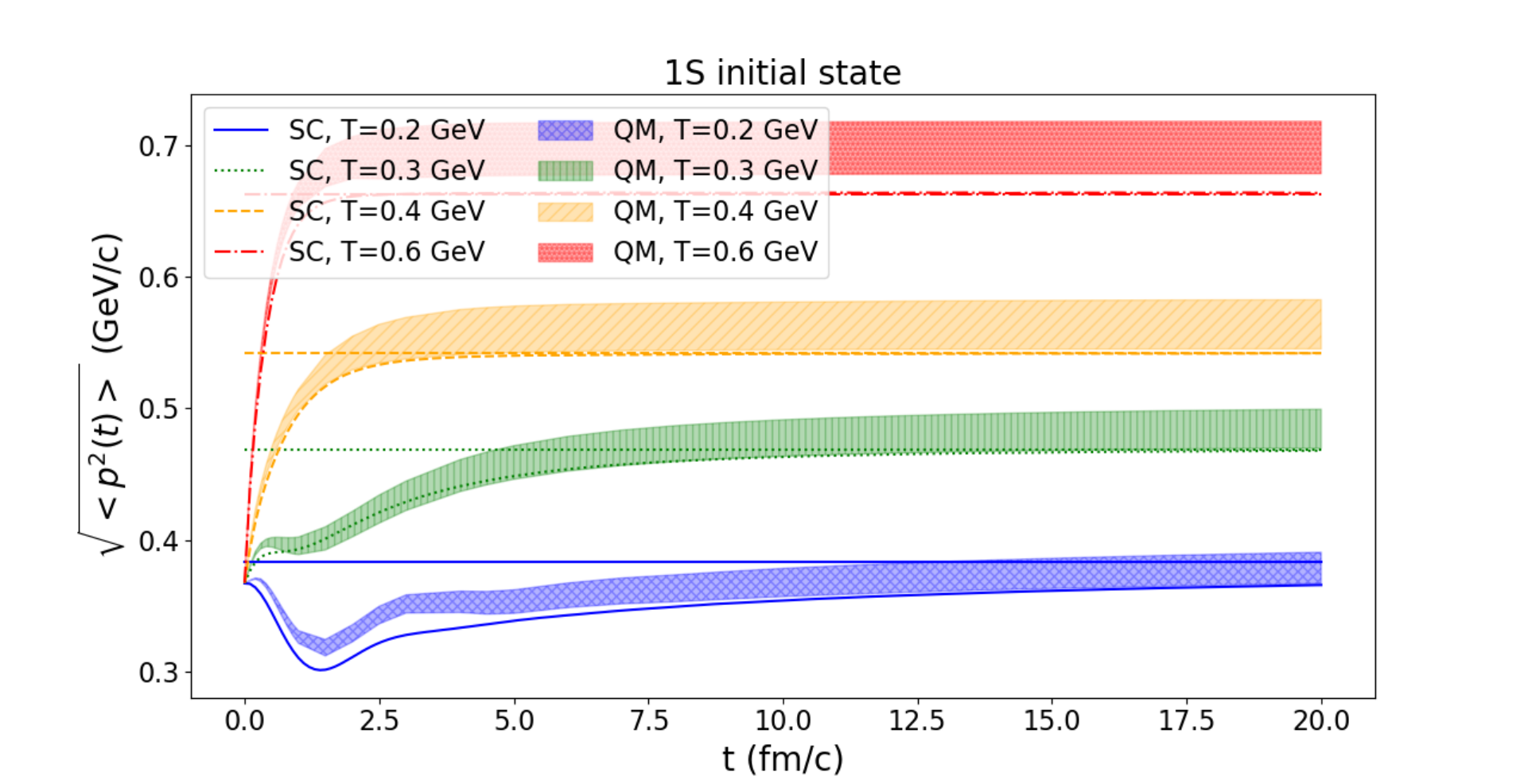}
	\caption{QM vs SC time evolution of root mean squared momentum for various QGP temperatures (the lower edges of the QM bands correspond to the calculation without $\mathcal{L}_4$). The horizontal lines refer to the stationary values expected from Gibbs-Boltzmann distribution, i.e.  $\sqrt{\langle p^2\rangle}(t=\infty)=\sqrt{MT/2}$. }
	\label{fig:sqrt-p-squared}
\end{figure}

\begin{figure}
	\centering
	\includegraphics[width=0.5\textwidth]{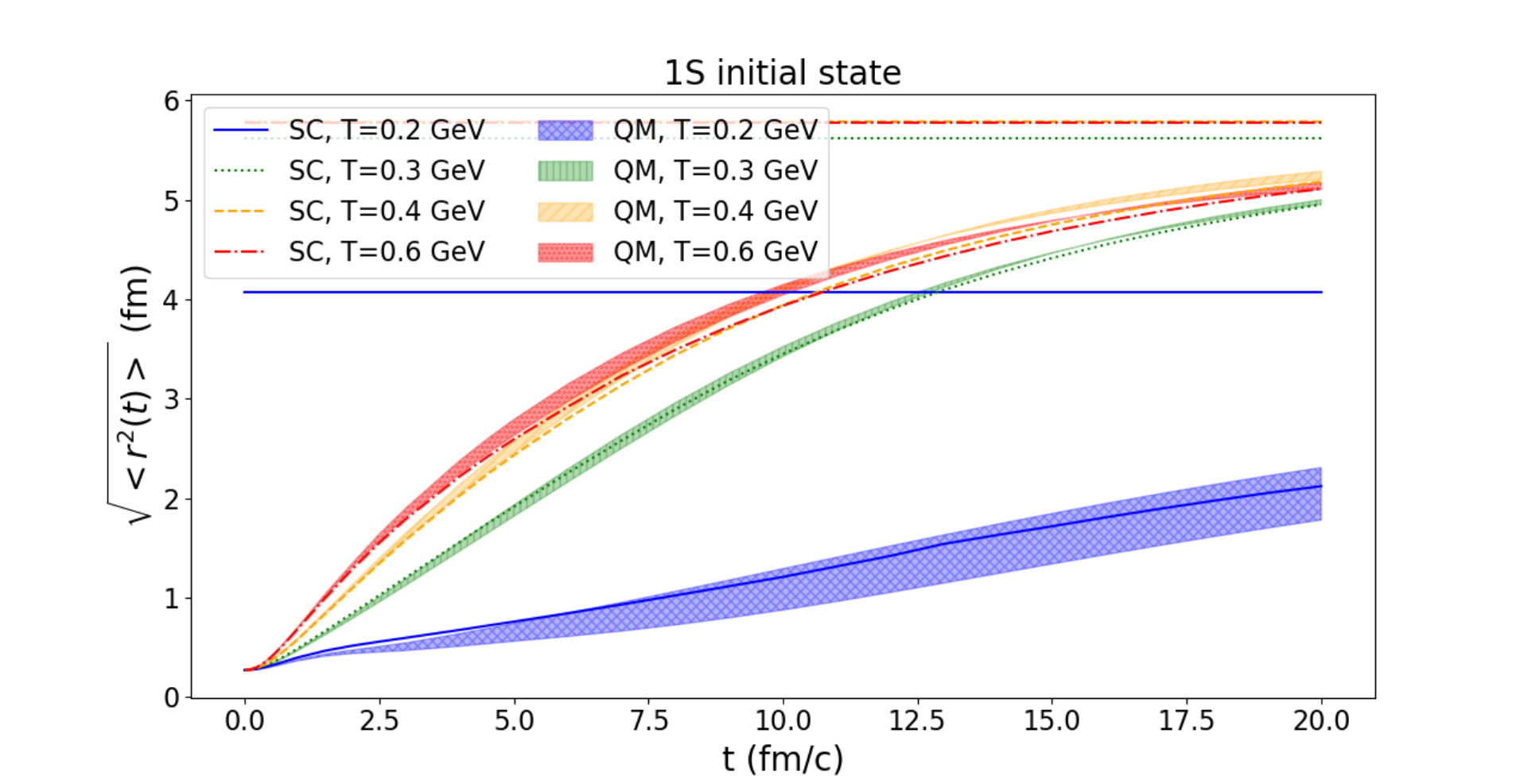}
	\caption{QM vs SC time evolution of root mean squared radius for various QGP temperatures. The horizontal lines refer to the stationary values expected from Gibbs-Boltzmann distribution. These  indeed depend on the temperature since the density is more localized toward small $r$ for low temperature, due to the attractive potential. }
	\label{fig:sqrt-r-squared}
\end{figure}

To gain a more quantitative understanding on the thermalization time scales in both the momentum and position sectors of phase space as well as on the corresponding asymptotic steady states,
we present in Figs. \ref{fig:sqrt-p-squared} and \ref{fig:sqrt-r-squared} the time evolution of the root mean squares of the relative momentum and of the relative distance of the heavy quarks. 

We observe in Fig.~\ref{fig:sqrt-p-squared}  a fast thermalization of the momentum sector, which is rather well reproduced by the SC description. Let us  
recall that, in the semiclassical picture,  the rate of increase of $\langle p^2 \rangle$ due to collisions is controlled by  $\eta(0)=\tilde{W}''(0)$ (see Eq.~(\ref{eq:defcoeff})).\footnote{Small size bound states are probing the imaginary potential around origin, hence $\eta(0)$.}  For our imaginary potential, this results in $\eta(0)=0.095 , 0.286 , 0.581$, and 1.53 $\mathrm{GeV}^2/\mathrm{fm}$, respectively for $T=0.2, 0.3, 0.4$, and $0.6$~GeV, showing a fast increase with $T$ ($\approx  T^{2.5}$), as observed on Fig.~\ref{fig:sqrt-p-squared}, leading to a much faster thermalization with larger $T$. At the lowest temperature the effect of the collisions is masked at early time  by the unitary evolution.\footnote{The transient drop of $\langle p^2 \rangle$ observed until 1.5 fm/$c$ for the lowest temperature is due to the conversion of kinetic energy into potential energy: As the initial state is slightly more compact than the first eigenstate of the in-medium Hamiltonian, it indeed starts by expanding under the slight excess of quantum pressure.} 

Focusing on the ``late time" momentum thermalization, we notice that the quantum description slightly overshoots the Boltzmann asymptotic value, the more so the higher the temperature, the effect being amplified  when ${\cal L}_4$ is included. This indicates that the steady states in the two descriptions are not identical. Indeed, while the semiclassical description exactly relaxes into a Gibbs-Boltzmann distribution, with $\langle p^2\rangle = \frac{M T}{2}$, the steady state solution of the quantum master equation,  in the quantum Brownian regime,  deviates from this distribution. Such a feature was already observed and discussed in \cite{Delorme:2024rdo}, where it was established that the variance of the relative momentum $p$ for the stationary distribution writes ${\rm var}(p)=\frac{MT}{2+\gamma}$, with $\gamma = \frac{\tilde{W}^{(4)}(0)}{16 MT \tilde{W}^{''}(0) }$, where $\tilde{W}^{(4)}(0)$ stems from the $\mathcal{L}_4$ operator. Schematically, this can be interpreted as if such term, mandatory for ensuring positivity  induces some deviation for ${\rm var}(p)$. More generally,  the deviation of the stationary solution of the Lindblad equation from a thermal state remains an issue,  see e.g. \cite{tupkary2022fundamental,Timofeev:2022tbl} and  references therein for recent discussions.

Interestingly, Fig.~\ref{fig:sqrt-p-squared} suggests that the omission of the operator $\mathcal{L}_4$ yields results closer to the Boltzmann-Gibbs distribution, and hence closer to the semiclassical result. 

In the position sector, thermalization appears to proceed in two steps:$\langle r^2 \rangle$ first grows $\propto \frac{\eta(0)}{M^2}\times t^3$, driven by the increase of $\langle p^2 \rangle$ (when the temperature is sufficiently high). When thermalization is reached in the momentum space, $\langle r^2 \rangle$ continues to grow $\propto D_s t$, with the spatial diffusion coefficient $D_s \sim \frac{T^2}{\eta(0)}$, implying a slightly faster growth for smaller temperatures (as can be observed by comparing the three curves corresponding to $T=0.3,0.4$ and 0.6 GeV for $t > 5\,\mathrm{fm}/c$). Full thermalization is therefore not reached before $t \sim \frac{L^2}{D_s}$ corresponding to the full delocalization of the heavy quarks inside the box. Thus thermalization in momentum and in space sectors generically proceed with different time scales, the 2nd being larger than the first as illustrated on Fig.~\ref{fig:sqrt-r-squared} where space-thermalization is still not observed for $t=20\,\mathrm{fm}/c$. Note that the thermalization time in the position sector depends on the size of the box used in the numerical simulations.

\subsection{Decoherence and classicalization}
\label{sec_Decoherence and Classicalization}
The structures observed  in Fig.~\ref{fig:snapshots-wigner} in the quantum Wigner function at $t=1$ $\mathrm{fm}/c$,  is suggestive of interferences which are absent in the semiclassical density matrix. Given that the initial 1S vacuum state is not an eigenstate of the in-medium potential, the evolution induced by the in-medium potential generates an interference between the 1S vacuum state and, for instance, the 2S excited state. This interference  manifests  itself through  negative contributions to the Wigner function, spoiling the potential interpretation of the latter as a classical phase-space distribution function. However, as soon as the non-unitary dynamics take over, the decoherence  quenches interferences, the Wigner function stays positive and a better agreement with the semiclassical description is observed in Fig.~\ref{fig:snapshots-wigner}. 

\begin{widetext}
	
	\begin{figure}[H]
		\centering
		\includegraphics[width=0.49\textwidth]{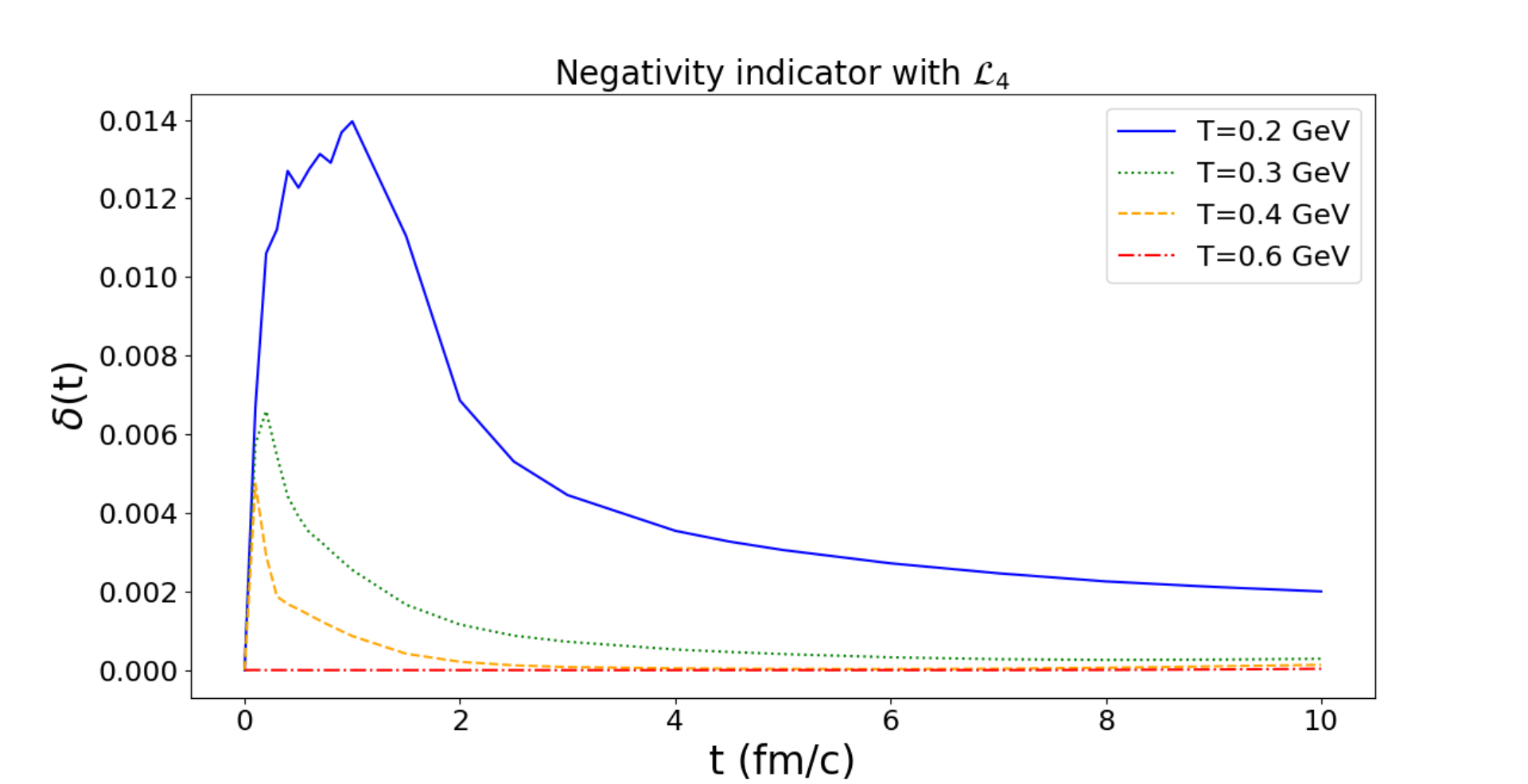}
		\includegraphics[width=0.49\textwidth]{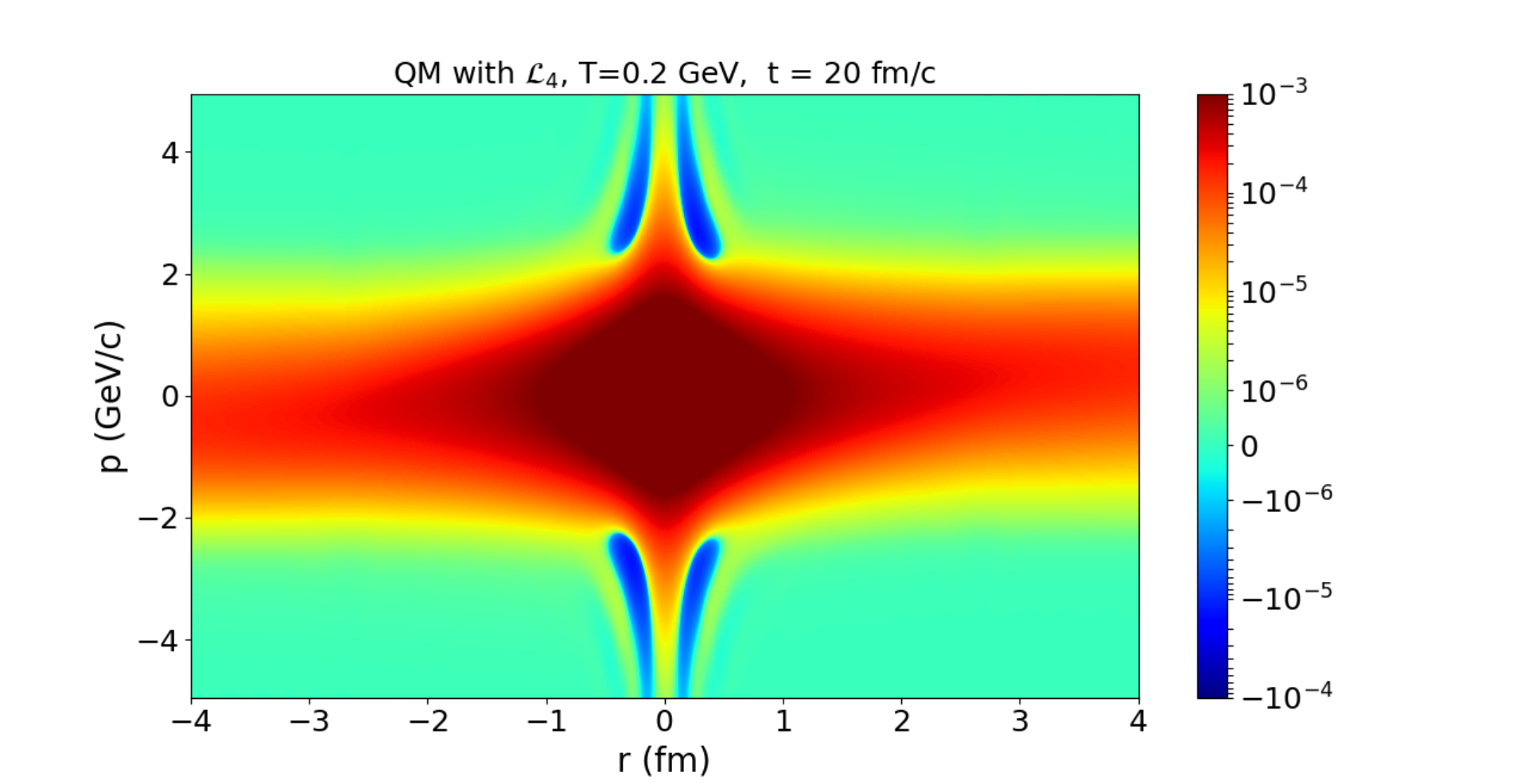}
		\caption{Left: The time evolution of the negativity indicator $\delta(t)$ for the case a quantum dynamics with $\mathcal{L}_4$ included in the dynamics. Right: the corresponding Wigner distribution at $t=20\,{\rm fm}/c$ for $T=0.2$ GeV. Small negative values can be detected at very small $r$ and large $p$. }
		\label{fig:Wigner-negativity}
	\end{figure}
	
\end{widetext}

In order to quantify the positivity of the Wigner function, we use the following  indicator, introduced in \cite{kenfack2004negativity}:
\begin{eqnarray}
	\delta\left(t\right)&=&\frac{1}{2\pi \hbar}\int_{{r,p}}\left(\left|\mathcal{D}\left(r,p,t\right)\right|-\mathcal{D}\left(r,p,t\right)\right)\nonumber\\&=&\frac{1}{2\pi \hbar}\int_{{r,p}}\left|\mathcal{D}\left(r,p,t\right)\right|-1.
\end{eqnarray}
This vanishes iff $\mathcal{D}>0$ everywhere in phase-space, so that a positive value of $\delta$ signals the existence of phase-space regions where the Wigner function is negative. The generic behavior of  $\delta(t)$ is shown in Fig.~\ref{fig:Wigner-negativity}: After a rapid increase at small time, it reaches  a maximum value and then decreases to a value close to 0 when the temperature is high enough. The maximum is a decreasing function of increasing  temperature, and occurs when the dynamics starts to be dominantly non-unitary.  For $T=$0.6 GeV, the in-medium potential vanishes, so that there is no source of interference, and  $\delta(t)=0$ at all times. For $T=0.2$ GeV, $\delta(t)$  decreases very slowly at later times: this is due to the persistence at late time  of a bound state responsible for interferences. These affect the Wigner function at small $r$ and large $p$, as can be seen from the right panel of Fig. \ref{fig:Wigner-negativity}. Overall, however, $\delta(t)$  remains small as compared to the norm of $\mathcal{D}$.

\subsection{Semiclassical expansion}
\begin{figure}[H]
	\centering
	\includegraphics[width=0.5\textwidth]{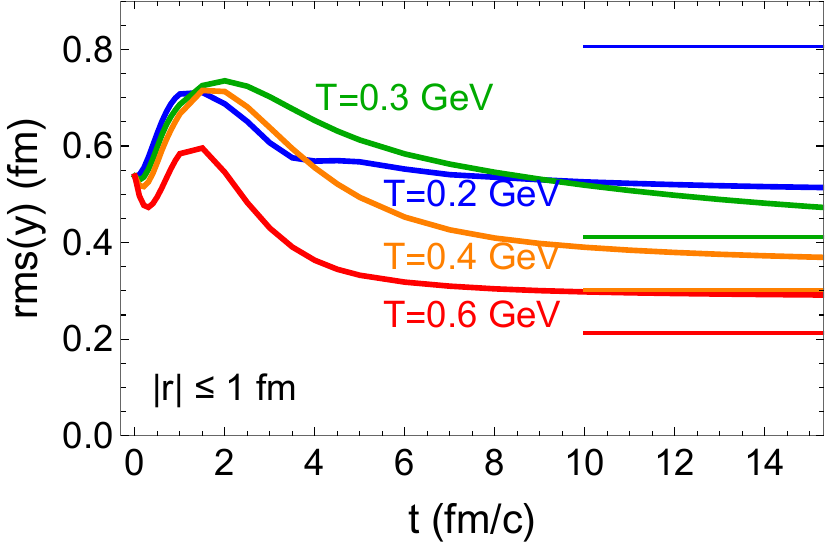}
	\caption{Time evolution of the root mean squared $\sqrt{\langle y^2\rangle}$ (thick curves),  compared to $y_\mathrm{infl}$, the inflection point of the imaginary potential (thin horizontal lines, corresponding, from top to bottom, to $T$=0.2, 0.3, 0.4 and 0.6, in GeV).}
	\label{fig:coherence-length-vs-turning-point-new }
\end{figure}

Further insight can be gained by examining the non-diagonal element ${\cal D}(s,s')$ of the density matrix. We expect indeed the semiclassical approximation to be valid when $ \sqrt{\langle y^2 \rangle}$, where $y=s-s'$, is small enough. As a rough touchstone for $ \sqrt{\langle y^2 \rangle}$, we take the inflection point $y_{\rm infl}$ of the imaginary potential (which depends on the temperature), below which $\tilde{W}$ can be well  approximated by a quadratic function, and the semiclassical approximation could be expected to be reasonably accurate. The time evolution of $ \sqrt{\langle y^2 \rangle}$, obtained through a quantum calculation, and defined as $\langle y^2 \rangle=\int_{|r<1|} {\rm d}r {\rm d}y |\mathcal D| y^2 /\int_{|r<1|} {\rm d}r {\rm d}y |\mathcal D| $  is shown in Fig.~\ref{fig:coherence-length-vs-turning-point-new }, where $|r|\le 1\,{\rm fm}$ was imposed in the average in order to probe the region most relevant for quarkonia observables.

At early time and small temperature, the dynamics is essentially unitary. The increase of $\sqrt{\langle y^2\rangle}$ is related to the initial increase of the size of the initial vacuum state, which is smaller that the corresponding in-medium state, as already mentioned. As time increases, the behavior changes, reflecting the fact that for $T\gtrsim 0.3$ GeV, the early dynamics gradually becomes  dominantly  non-unitary. 
The decrease of $\sqrt{\langle y^2\rangle}$ is hence more pronounced, as the collisions play an increasing role, which one can relate to the fast increase of $\sqrt{\langle p^2\rangle}$ observed in Fig.~\ref{fig:sqrt-p-squared} for $T\ge 0.4$ GeV.  At later times, the non-unitary dynamics drives the system to a stationary state, and $\sqrt{\langle y^2\rangle}$ evolves toward the thermal wavelength $\sqrt{2/M T}$.  The condition $\sqrt{\langle y^2\rangle} \lesssim y_\mathrm{infl}$ is approximately fulfilled in all cases, except at the highest temperature considered, where $y_{\rm infl}$ appears to be  much smaller than the initial size of the wave packet. However, this does not entail a breakdown of the semiclassical approximation: when evaluated with the SC approximation, $\sqrt{\langle y^2\rangle}$ exhibits a dependence with time quite close to that displayed in Fig.~\ref{fig:coherence-length-vs-turning-point-new } for the quantum evaluation. 

\begin{widetext}
	
	\begin{figure}[H]
		\centering
		\includegraphics[width=0.49\textwidth]{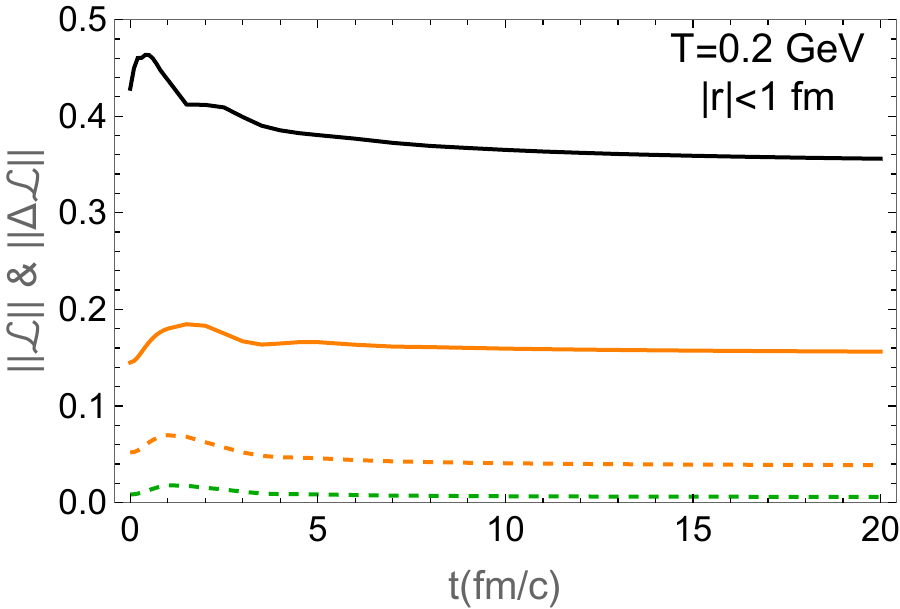}
		\includegraphics[width=0.49\textwidth]{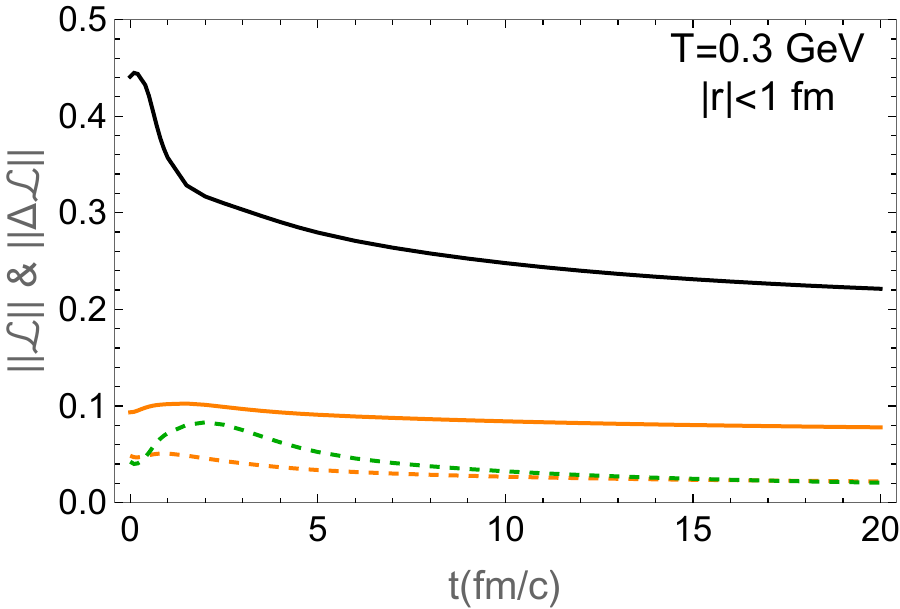}
		\includegraphics[width=0.49\textwidth]{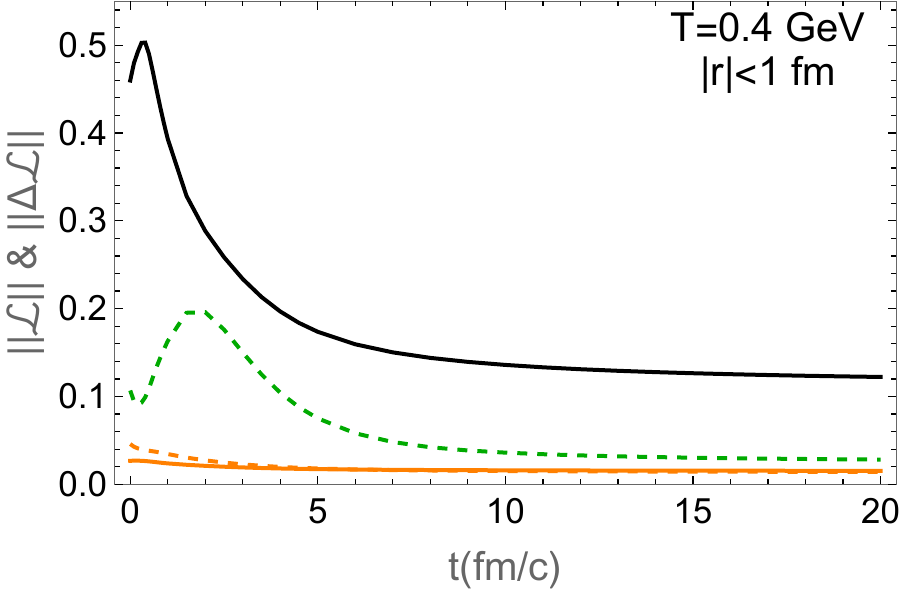}
		\includegraphics[width=0.49\textwidth]{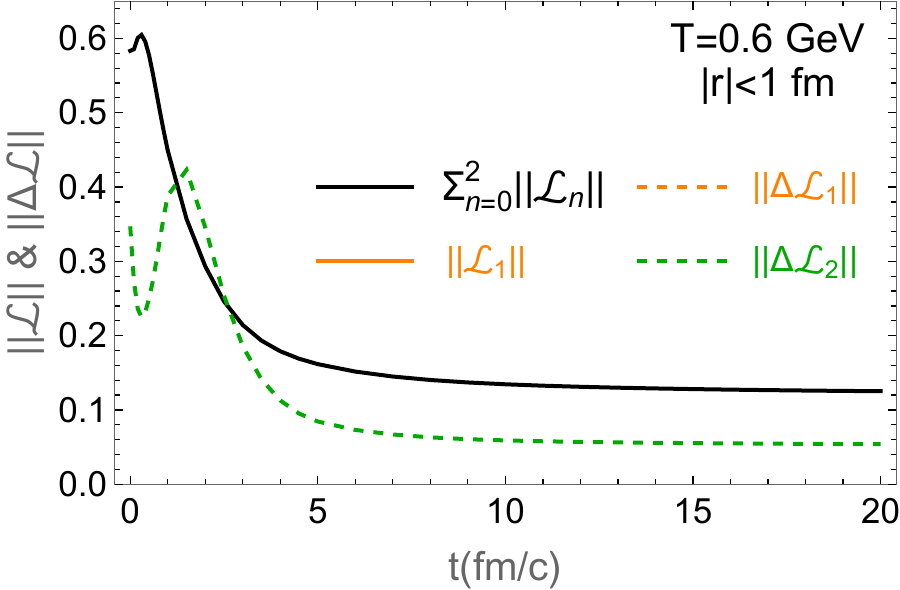}
		\caption{Time evolution of the ``norm" $\|\mathcal{L}\|$ and of both errors $\| \Delta \mathcal{L}_1\|$ and $\| \Delta \mathcal{L}_2\|$ for various QGP temperatures; see text for definitions. For $T=0.6$ GeV, one has $V=0$  implying $\| \Delta \mathcal{L}_1\|=\| \mathcal{L}_1\|=0$. }
		\label{fig:abs errors-new}
	\end{figure}
\end{widetext}

To explore this feature further, we plot on Fig.~\ref{fig:abs errors-new}, the quantities $\|\Delta \mathcal{L}_1\|$ and $\|\Delta \mathcal{L}_2\|$  which measure the deviations of the SC calculation from the quantum one, via a suitable average of the operators $\mathcal{L}_1$ and $\mathcal{L}_2$.

These quantities are defined as 
\begin{equation}
	\| \Delta \mathcal{L}_i\| = \frac{\int_{|r|<1} {\rm d}r {\rm d}y |(\mathcal{L}_i- \mathcal{L}_i^{\rm SC}) \mathcal{D}| }{\int_{|r|<1} {\rm d}r {\rm d}y |\mathcal{D}| },
\end{equation}
and we also introduce, as an overall reference $\| \mathcal{L}\| = \sum_{n=0}^{4}\| \mathcal{L}_n\| \approx 
\sum_{n=0}^{2}\| \mathcal{L}_n\|$ with
\begin{equation}
	\| \mathcal{L}_n\| = \frac{\int_{|r|<1} {\rm d}r {\rm d}y |\mathcal{L}_n \mathcal{D}| }{\int_{|r|<1} {\rm d}r {\rm d}y |\mathcal{D}| }.
\end{equation}
Here $\mathcal{D}$ is chosen as the solution of the QME, and  the denominator ensures proper normalization. The restriction $|r|<1$ has been imposed as was done earlier when defining $\langle y^2\rangle$. 

Since $(\mathcal{L}_2- \mathcal{L}_2^{\rm SC}) \propto -\frac{\tilde{W}^{(4)}(0)+\tilde{W}^{(4)}(r)}{192}\,y^4$ for a regular imaginary potential $\tilde{W}$ (see Appendix A), the behavior of $\| \Delta \mathcal{L}_2\|$ observed on Fig.~\ref{fig:abs errors-new} reflects that of the rms($y$) displayed on Fig.~\ref{fig:coherence-length-vs-turning-point-new }. Up to $t\approx 3\,{\rm fm}/c$, it becomes increasingly large with increasing $T$, and can even become of a comparable magnitude as $ \|\mathcal{L}\|$ for the two highest temperatures considered. For later times, however, it drops  in a similar way as $\langle y^2 \rangle$ and becomes eventually quite moderate. 

We turn now to the unitary evolution that dominates at early time and low temperature. The quantities $\|\Delta \mathcal{L}_1\|$, $\mathcal{L}_1$ and $\mathcal{L}_1^{\rm SC}$ are evaluated respectively  with the ``native" 1D real potential and the regularized one (see appendix \ref{sec:regpot})\footnote{Note that whether one uses the native or the regularized potential has little impact on the quantum evolution.}.

As one can see from Fig.~\ref{fig:abs errors-new}, the contribution from the real potential is only significant for the 2 lowest temperatures. In this range, $\|\Delta \mathcal{L}_1\|$ is at most 50\% of $\|\mathcal{L}_1\|$  and at most $10\%$ of $\|\mathcal{L}\|$ for early times. The largest deviations occur at early time, and as we shall verify now, at small relative distance. Let us define  the following $r$-dependent quantities
\begin{equation}
	| \mathcal{L}_1^{{\rm SC }}| = \frac{\int {\rm d}y |\mathcal{L}^{{\rm SC }}_1 \mathcal{D}| }{\int {\rm d}y |\mathcal{D}| }
	\quad \text{and} \quad 
	|\Delta \mathcal{L}_1| = \frac{\int {\rm d}y |(\mathcal{L}_1- \mathcal{L}_1^{\rm SC}) \mathcal{D}| }{\int {\rm d}y |\mathcal{D}| }.
\end{equation}

\begin{figure}
	\centering
	\includegraphics[width=0.5\textwidth]{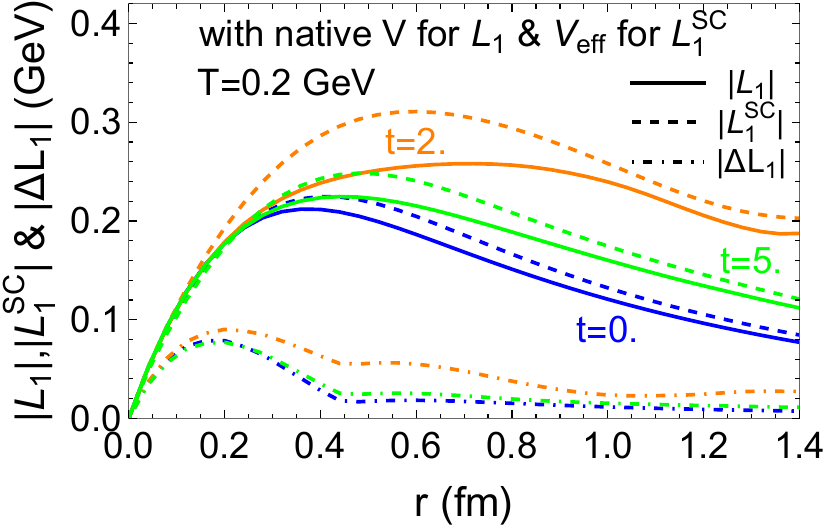}
	\caption{ $|\mathcal{L}_1|$ (plain lines), $|\mathcal{L}_1^\mathrm{SC}|$ (dashed lines) and $|\Delta \mathcal{L}_1|$ (dot-dashed lines) at $t=0$, 2, and 5 fm/$c$, for an evolution in a QGP at $T=0.2$ GeV.
		See text for definitions. }
	\label{fig:errorL1loc-new}
\end{figure}

These are plotted in Fig.~\ref{fig:errorL1loc-new}, for $T=0.2\,{\rm GeV}$. 
For the three values of  times under investigation, 
$ |\mathcal{L}_1^{{\rm SC}}|\approx  |\mathcal{L}_1|$ on the $r$-range considered, with the deviation $|\Delta \mathcal{L}_1|$ being more concentrated 
at small $r$. This is not surprising, since  this is where the corrections associated with the higher derivatives of the real potential are expected to be the most significant in the Wigner-Moyal expansion.\footnote{One should remember that $|\mathcal{L}_1^{{\rm SC }}| =  |\mathcal{L}_1| \Rightarrow \!\!\!\!\!\!\! / \;\;\; |\Delta \mathcal{L}_1|=0$.}

\subsection{Distances}
\label{section_quantitative}
Various kinds of distances are often used to quantify the differences between solutions of equation of motion. Here we complete the comparison between the quantum and the semiclassical evolutions with the help of the following distance between the Wigner transforms of the corresponding density matrices \footnote{The so-called Schatten norm is often  used in the field of quantum information, which corresponds to a trace distance $\sqrt{\int\text{d}r\text{d}p\,\left(\mathcal{D}_\mathrm{QM}\left(t,r,p\right)-\mathcal{D}_\mathrm{SC}\left(t,r,p\right)\right)^{2}}$ \cite{Lidar:2019qog}. This distance was utilized in \cite{daddi2024} to assess the validity of the semiclassical description of in-QGP quarkonia dynamics. Here we 
	privilege the more elementary definition (\ref{eq:trace-distance-def-2}).}  
\begin{equation}
	d\left(t\right)= \int\frac{\text{d}r\text{d}p}{2\pi \hbar}\left|\mathcal{D}_\mathrm{QM}\left(t,r,p\right)-\mathcal{D}_\mathrm{SC}\left(t,r,p\right)\right| 
	\label{eq:trace-distance-def-2}.
\end{equation}
This considers the $m\times n$  (Wigner) density matrix as a vector of length $m*n$ and then uses the usual $\|\cdot\|_1$ vector norm. 

Contrarily to alternate definitions, this distance is not bounded for arbitrary distributions. However, as our Wigner distributions are mostly positive (see the discussion on the $\delta$ indicator in the previous subsection), one has $\int\frac{\text{d}r\text{d}p}{2\pi \hbar}\left|\mathcal{D}\left(t,r,p\right)\right|\approx 1$, which thus constitutes a natural reference for the magnitude of  $d$. 

\begin{figure}[H]
	\centering
	\includegraphics[width=0.53\textwidth]{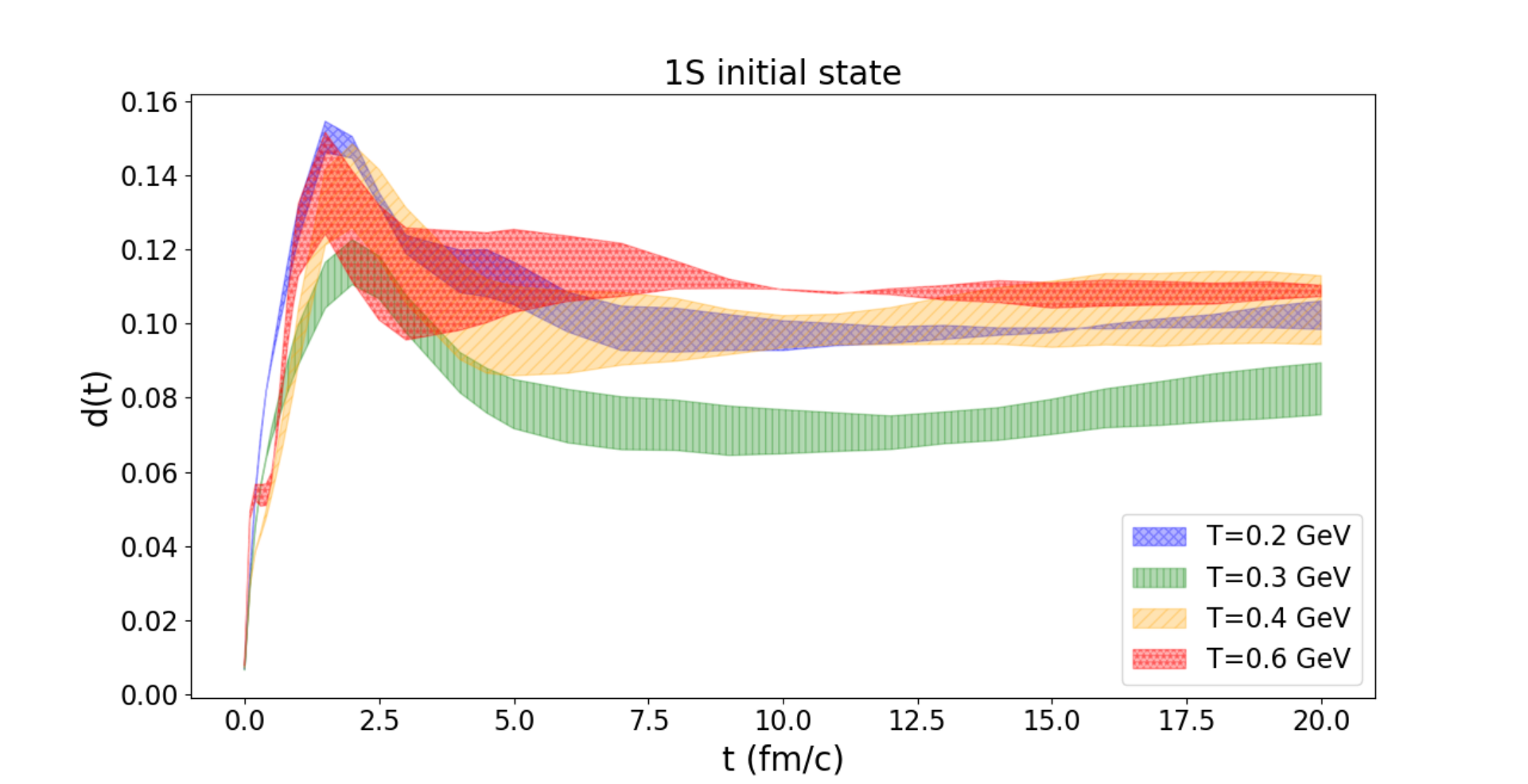}
	\caption{ Time evolution of the distance $d(t)$    for various QGP temperatures. }
	\label{fig:distance}
\end{figure}

In Fig.~\ref{fig:distance}, we present the time evolution of the distance $d$. Evidently, since  the same initial state is used in both descriptions, one expects $d\left(t=0\right)=0$.\footnote{A small deviation  ($\lesssim 5\times 10^{-3}$) actually occurs : it is due to the fact that a finite set is used in the  stochastic sampling of the initial Wigner function when solving the Langevin equation.} The distance exhibits the same time evolution pattern for the various QGP temperatures.
A maximum distance $d$ of $\sim 12-14 \%$ is reached at early time $t\lesssim 2 \text{fm/c}$ for all temperatures, and then saturates to a value that depends weakly on the temperature.

\begin{figure}
	\centering
	\includegraphics[width=0.53\textwidth]{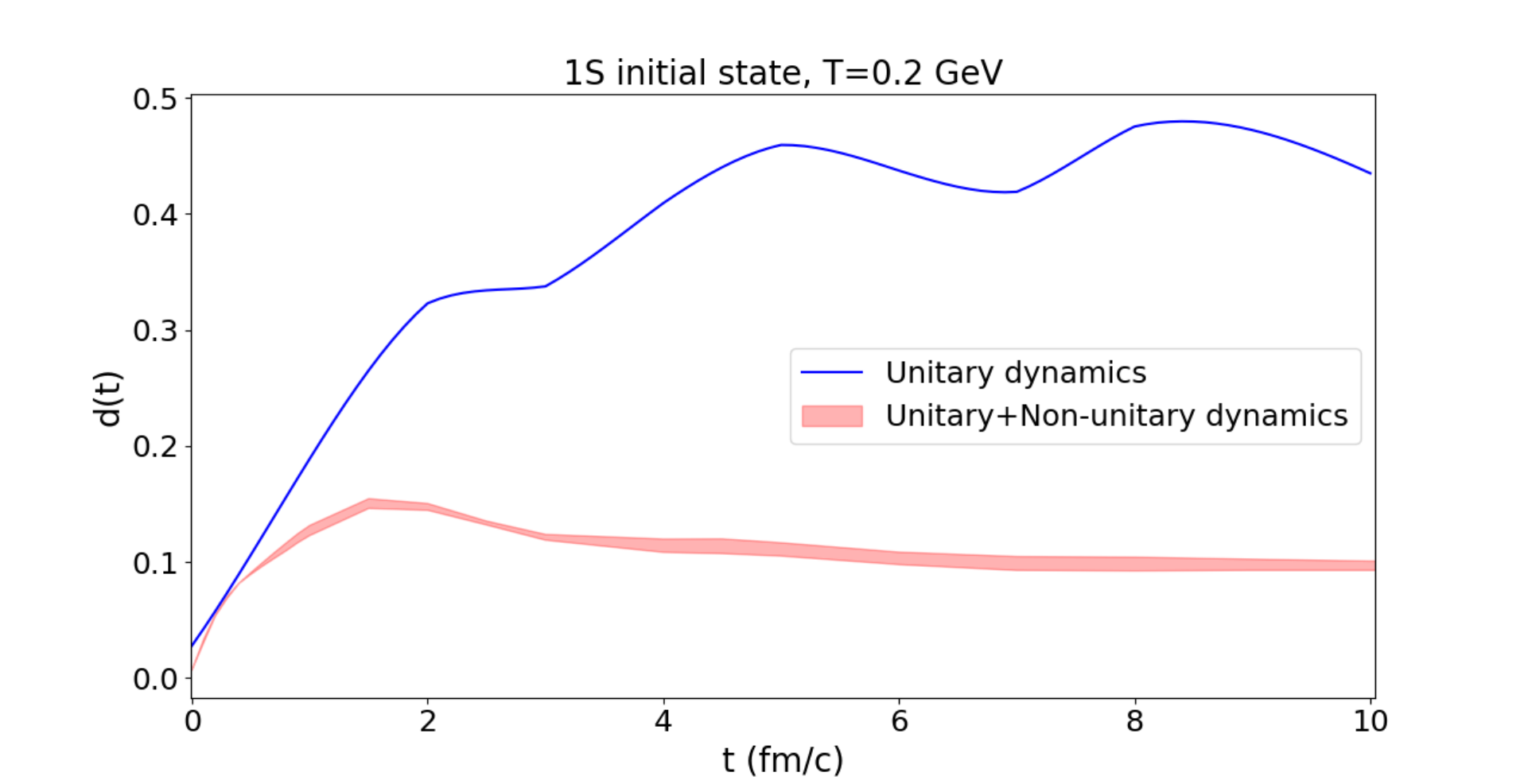}
	\caption{ Comparison of the distance  $d(t)$  time evolution in presence and absence of the non-unitary component of the dynamics,  for $T=0.2$ GeV. } 
	\label{fig:trace-dist-L1}
\end{figure}

At early times,  the quantum  features  of the dynamics, e.g. the interference observed in Fig.~\ref{fig:snapshots-wigner}, are not well captured by the semiclassical description. This induces some discrepancy between the quantum and semiclassical descriptions which is quantified by the increase of $d(t)$ in Fig.~\ref{fig:distance} at early time. This increase persists  until quantum decoherence takes over, as discussed in the previous subsection. In order to confirm the role of this mechanism, a comparison is presented in Fig.~\ref{fig:trace-dist-L1} for $d(t)$ calculated respectively in the absence and in the presence of the non-unitary component of the dynamics. When the  dynamics are unitary,  the distance increases continuously, and ultimately oscillates at late times around a high value. In contrast, the non-unitary evolution that results from the coupling with the environment  tames the early increase of $d(t)$. At later times, one observes on Figs.~\ref{fig:distance} and \ref{fig:trace-dist-L1} that the distance stabilizes to an approximately constant positive value. This non vanishing value of the distance can, to a large extent, be attributed  to the fact that the QME and the SC equations do not lead to the same asymptotic state, as already emphasized (the SC evolution converges to the Boltzmann-Gibbs distribution with a temperature equal to that of the heat bath while, as discussed in \cite{Delorme:2024rdo},  the effective temperature attached to the solution of the Lindblad equation is slightly larger than that of the heat bath, irrespective of the QGP temperature). 

A more detailed analysis of the behavior of $d(t)$ in terms of marginal distances $d_r$ and $d_p$  is presented in Appendix \ref{sec:ddrdp}. It reveals non-monotonous trends associated to the quantum corrections of both components of the complex potential, leading to a more subtle temperature-dependence than the decrease of $\delta$ discussed in section~\ref{sec_Decoherence and Classicalization}, while the temperature $T = 0.3$ GeV represents a (somewhat accidental) balance between reducing errors in the unitary and non-unitary components of the dynamics, thereby minimizing the distance $d$, as seen in Fig.~\ref{fig:distance}. 

\section{Quarkonium observables}
\label{local-comparaison}

In this section, we consider the evolutions of observables which are sensitive to the small distance dynamics of the quarkonia, that is we focus on spatial regions where the $c$ and the $\bar{c}$ are located within the range of the attractive potential.

\subsection{General features of the evolution of the 1S state}

\begin{widetext}
	
	\begin{figure}
		\centering
		\includegraphics[width=0.48\textwidth]{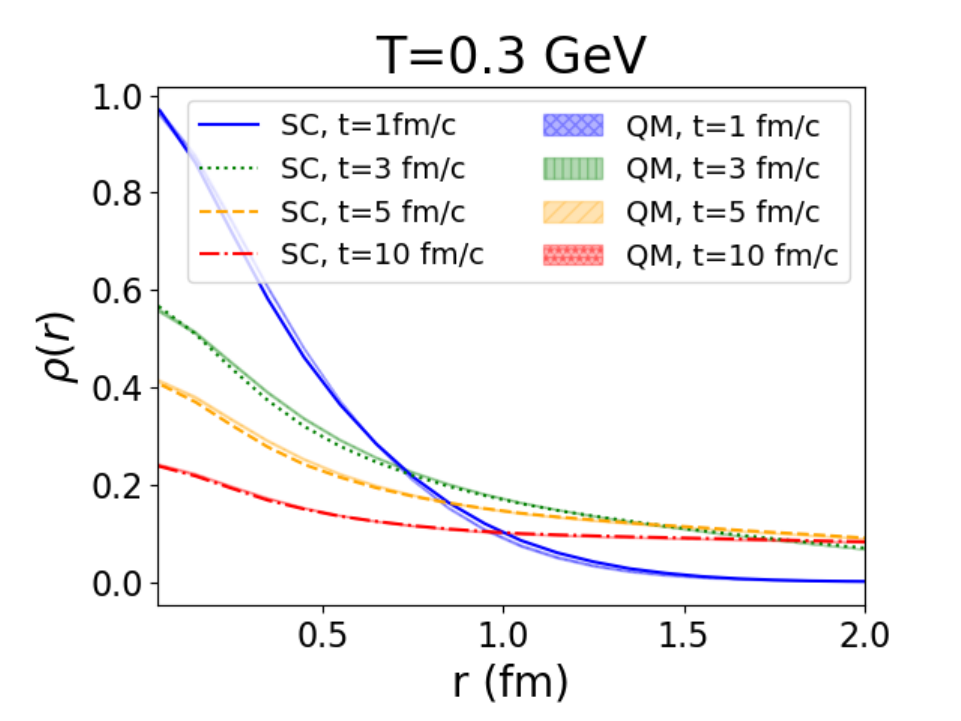}
		\includegraphics[width=0.48\textwidth]{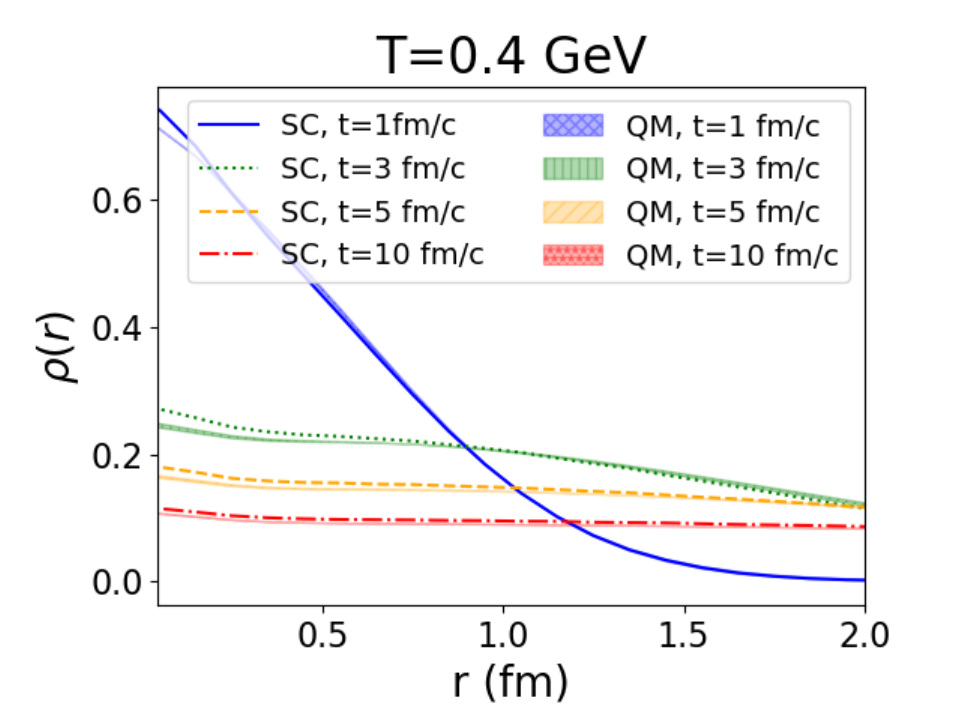} 
		\caption{QM vs. SC time evolution of the position probability density distribution resulting from the evolution of a 1S state. Left: T=0.3 GeV. Right: T=0.4 GeV.
			Note that the presence of the $\mathcal{L}_4$ in the QME evolution has no visible impact on the results. }
		\label{rho(r)-comparaison-0.3GeV}
	\end{figure}
	
\end{widetext}

We first consider, on Fig.~\ref{rho(r)-comparaison-0.3GeV}, the spatial density at small relative distance for various times and two representative QGP temperatures (0.3 and 0.4 GeV). For 0.3 GeV, the agreement between the quantum and SC evolution is impressive.  This is not unexpected, in view of the discussions in the previous section, where we saw that the deviations between the corresponding density matrices were by about ten percent, as seen on Fig.~\ref{fig:distance}.
For 0.4 GeV, the semiclassical evaluation of $\rho(r)$ exceeds slightly, at small $r$,  that of the quantum calculation, which can be related -- see Figs. \ref{fig:sqrt-p-squared}
and \ref{fig:sqrt-r-squared} -- to a (tiny) excess of the $\langle p^2\rangle$ and $\langle r^2\rangle$ growth for the quantum dynamics for $T$=0.4 and 0.6 GeV.

\begin{widetext}

	\begin{figure}[H]
		\centering
		\includegraphics[width=0.49\textwidth]{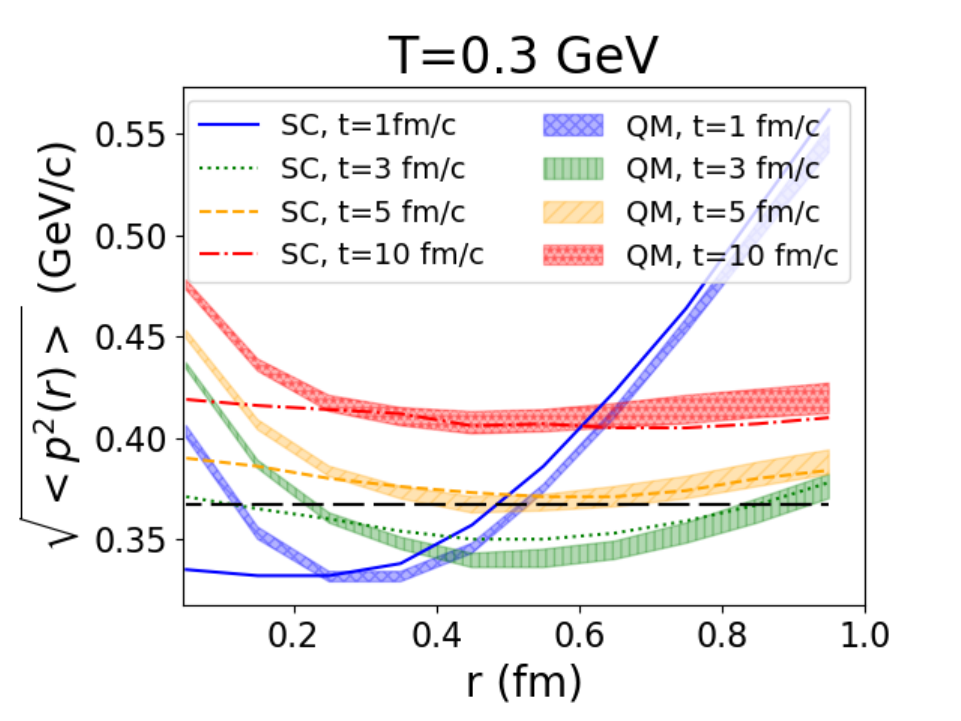}
		\includegraphics[width=0.49\textwidth]{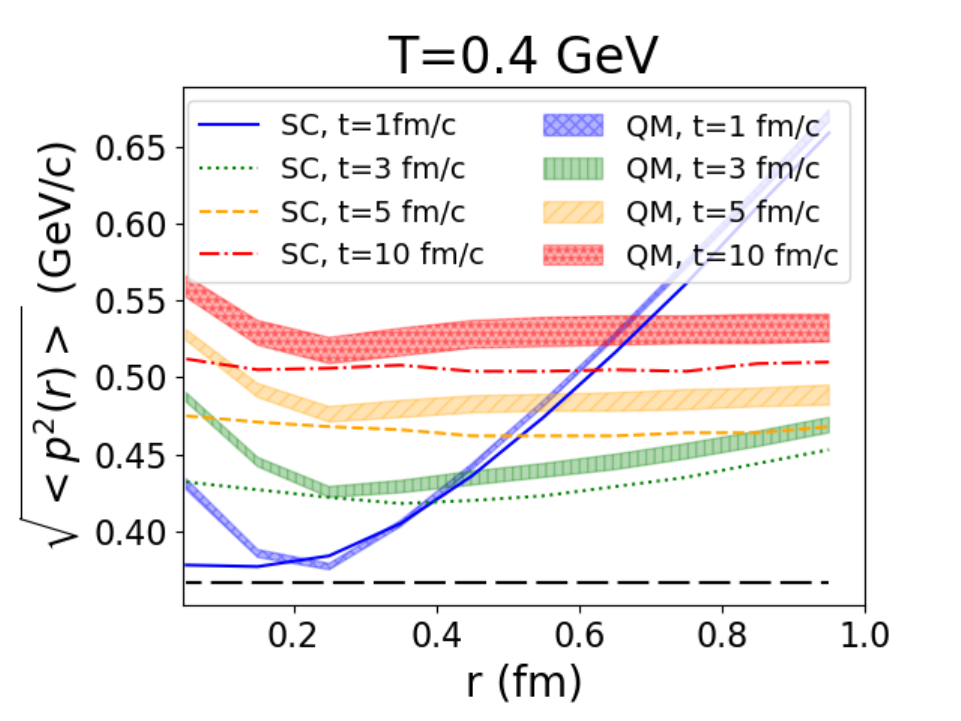}
		\caption{QM vs SC root mean squared momentum as function of position at different times. Left: T=0.3 GeV. Right: T=0.4 GeV. On both graphs, the horizontal dashed lines represent the initial mean squared momentum associated to the vacuum 1S state.}
		\label{fig:sqrt-p-squared-as-f(r)}
	\end{figure}

\end{widetext}

Further insight in this small $r$ region is obtained by looking at the root-mean-square momentum as a function of position, i.e $\sqrt{\langle p^2\rangle_r}$. which quantifies the agreement reached in momentum space at small $c\bar{c}$ distance. This is plotted  in Fig.~\ref{fig:sqrt-p-squared-as-f(r)}  for the same times and $T$ values as in Fig.~\ref{rho(r)-comparaison-0.3GeV}.
This comparison allows us to confirm the presence, for $T=0.4$ GeV (and beyond), of a systematic $\langle p^2\rangle$ excess of the quantum result at finite $r$ ($r \gtrsim 0.2-0.3\,{\rm fm}$), consistent with the excess already seen in the comparison between $r$-integrated $\langle p^2\rangle$ displayed on Fig.~\ref{fig:sqrt-p-squared}.

One observes also a clear but yet moderate deviation in the values of $\sqrt{\langle p^2\rangle}$ between the QM and SC calculations at short distance, which does not has the same (spurious) origin as the one at finite $r$.\footnote{A finer analysis reveals that a similar increase of $\sqrt{\langle p^2\rangle}$ at small $r$ is found in the Wigner transform of eigenstates for the real potential and hence in the asymptotic distribution which is dominated by these states. The raise of this structure at small-$r$ with time is therefore a direct consequence of the thermalization in the QM description, while no such structure is found in the SC case, as $\langle p^2\rangle_r$ is independent of $r$.} Such deviation between the QM and the SC descriptions would presumably be corrected by taking into account corrections in the Wigner-Moyal expansion of $\mathcal{L}_1$ in the SC calculation. As already identified in Fig.~\ref{fig:errorL1loc-new}, these corrections are indeed mostly located at small $r$.

\subsection{Quarkonia yields}
We come now to the  probability to observe a given charmonium (vacuum) state $\ket{\psi_i}$ after some evolution of the initial $c\bar{c}$ state during a time $t$. The corresponding probability $P_i$ is defined as 
\begin{equation}
	P_i(t)=\text{Tr}\left(\left|\psi_i\right\rangle\left\langle\psi_i\right|\hat{\mathcal{D}}_{c\bar{c}}(t)\right)=\int \frac{\text{d}r\text{d}p}{2\pi\hbar}\,\mathcal{D}_{i}(r,p)\mathcal{D}(r,p,t),
\end{equation}
where $\mathcal{D}_i$ and $\mathcal{D}$ are  the Wigner functions associated respectively to the stationary bound state ($\mathcal{D}_i$) and to the $c\bar{c}$ state ($\mathcal{D}(t)$). 

\begin{figure}[H]
	\centering
	\includegraphics[width=0.53\textwidth]{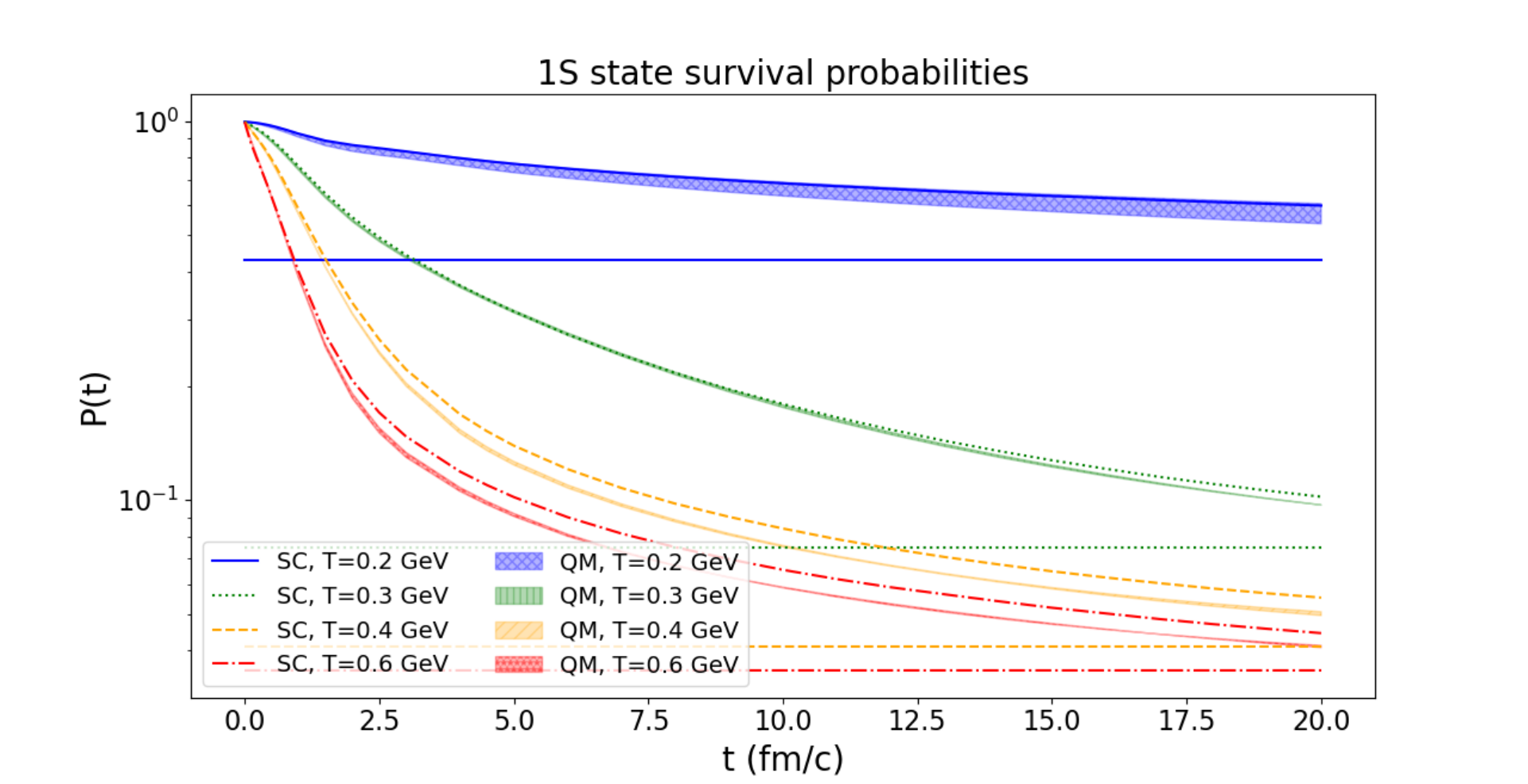}
	\caption{QM vs SC survival probability of the 1S state. The horizontal lines are the equilibrium values expected from the Gibbs-Boltzmann distribution.}
	\label{fig:1S-state-problt}
\end{figure}

We consider first the survival probability of the 1S initial state, i.e $\mathcal{D}_i=\mathcal{D}(t=0)=\mathcal{D}_{\text{1S}}$. This is shown in Fig.~\ref{fig:1S-state-problt} as a function of time.  Overall, the semiclassical description is in excellent  agreement with the quantum one, consistently with the very good agreement for both the spatial density and the momentum distribution at small $r$ observed in Figs.~10 and 11 respectively.\footnote{At late times and for $T=0.4$ GeV and beyond, the semiclassical description however tends to slightly overpredict the population of the ground state as compared to the QM calculation, by 10\% at the most. This is mostly the consequence of the small extra-depletion affecting the spatial density in the QM calculation observed on Fig. 10. } 

Fig.~\ref{fig:1S-state-problt} also reveals that survival probability does not reach its equilibrium value at $t=20$ fm/c. This may be attributed  to the slow thermalization of the position sector of phase space,  as already discussed in section \ref{section_global}. Indeed, upon solving the semiclassical equations up to 60 fm/c, it was observed that the equilibrium values of the survival probabilities were attained starting only from $t\simeq 50$ fm/c, see \cite{daddi2024}. This is in agreement with the thermalization time scales of $\sqrt{\langle r^2\rangle}$   observed in Fig.~\ref{fig:sqrt-r-squared}.    

We examine now the  probability to generate the 2S excited state, starting from the same 1S state as before. The  wave function of the 2S state is approximated by 
\begin{equation}
	\label{eq:2s-state-expression2}
	\psi_\mathrm{2S}\left(s\right)=Ae^{-s^2/B^2}+Cs^{2}e^{-s^2/D^2},
\end{equation}
with $A=0.741$ $\text{fm}^{-\frac{1}{2}}$, $B=0.22$ fm, $C=-3.4$ $\text{fm}^{-\frac{5}{2}}$, and $D=0.8$ fm. In contrast to the case of the 1S, which is associated to a positive Wigner transform,  the  Wigner transform of the 2S density matrix is negative in some regions of phase space, resulting in interference effects affecting the 2S yield. 

\begin{figure}[H]
	\centering
	\includegraphics[width=0.54\textwidth]{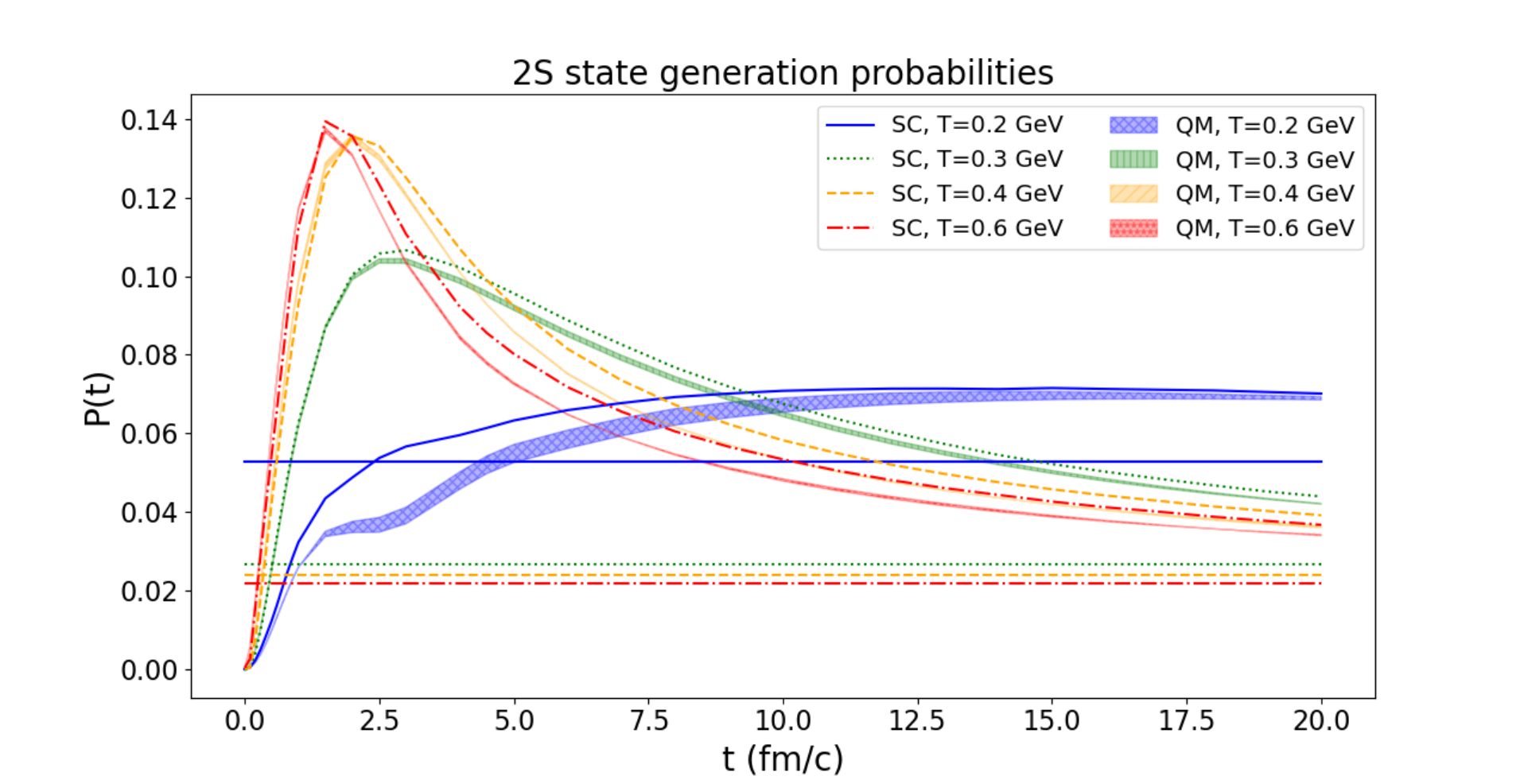}
	\caption{QM vs. SC generation probability of 2S $c\bar{c}$ state for various QGP temperatures.}
	\label{fig:2S-state-prob-hight-T}    
\end{figure}

The probability $P_{\text{2S}}$ is shown in Fig.~\ref{fig:2S-state-prob-hight-T}, as a function of time and for various QGP temperatures. The overall agreement between the quantum and the semiclassical calculations is of the same quality as observed earlier for the survival probability of the 1S state, with the semiclassical values slightly exceeding the quantum ones at late time, for the reason that has already been discussed. At the  lowest temperature,  interference effects manifest themselves by the slight oscillation of the quantum result at small time, which is not captured by the semiclassical calculation, see \cite{heller1976wigner}.  

However, the non-unitary evolution eventually suppresses these interferences, as shown in Fig.~\ref{fig:trace-dist-L1}. Consequently, the inaccuracies in the semiclassical approximation of the unitary dynamics become inconsequential as the non-unitary dynamics gradually takes over. And indeed, at high temperatures where the quantum decoherence is fast, we can see in Fig.~\ref{fig:2S-state-prob-hight-T} that the semiclassical description reproduces  very closely the results of the quantum description.

\subsection{Initial compact state}
\label{sec:compact}

We consider now, as a somewhat more realistic initial condition, a  compact state corresponding to the wave function given in Eq.~(\ref{eq:initial-state-def}) with $\sigma=0.165$~fm.  

\begin{figure}[H]
	\centering
	\includegraphics[width=0.54\textwidth]{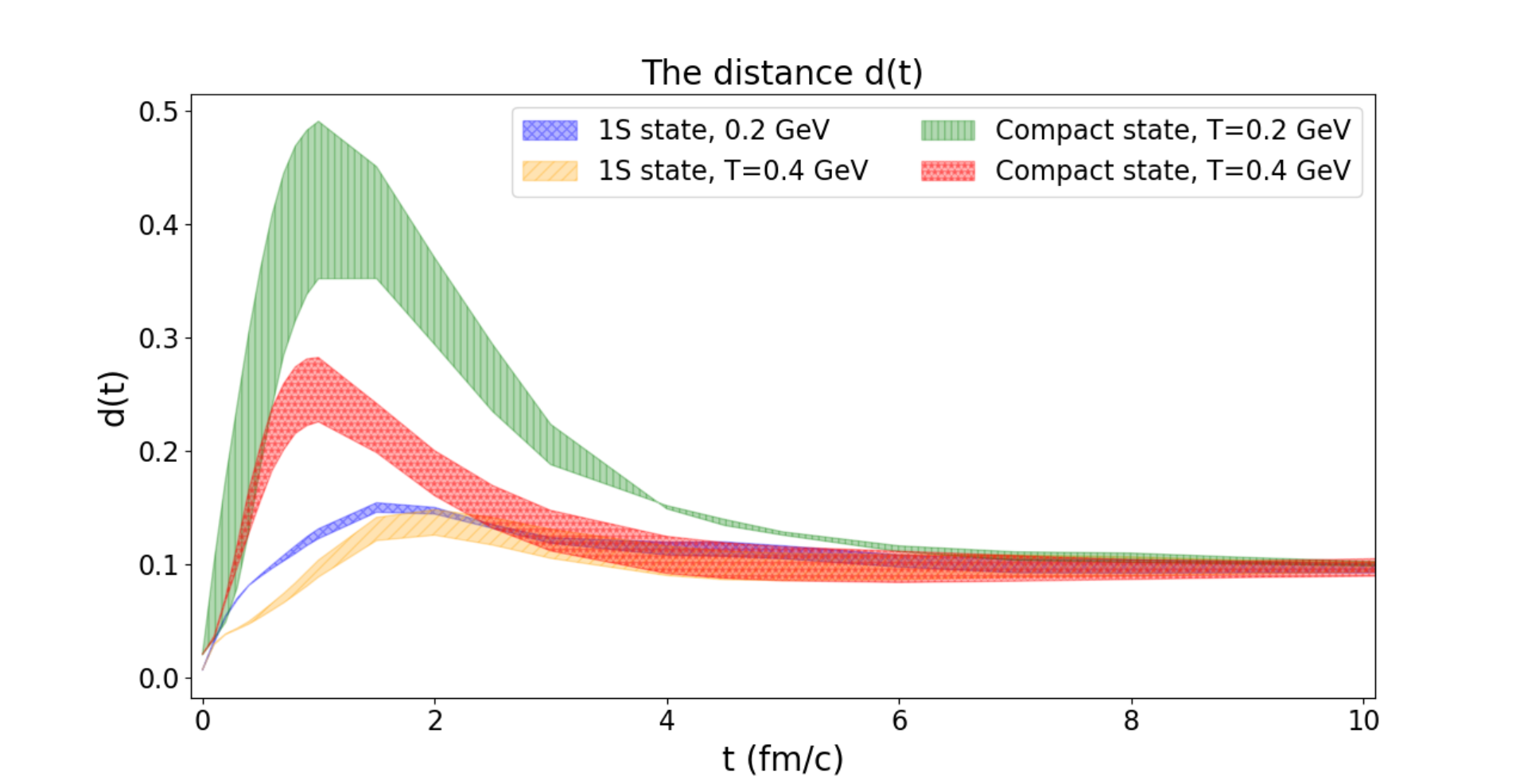}
	\caption{Comparison of the QM to SC distance $d(t)$  for different initial states and two values of $T$.   }
	\label{fig:trace-distance-compact-state}
\end{figure}

We first compare the evolution of this compact state with that of the 1S state studied earlier.  The distance $d(t)$ for the two initial states is shown in Fig.~\ref{fig:trace-distance-compact-state} for $T=0.2$ and 0.4~GeV. One sees that the distance grows to larger values for the compact initial state.
Starting from a compact state results indeed in a faster coherent expansion due to the larger quantum pressure. In the absence of coupling with the environment, the unitary dynamics would drive the density matrix $\mathcal{D}(s,s')$ toward regions of large values of $|y|$, where the SC approximation is less accurate, hence the increase of the distance. As discussed earlier, the growth of the distance is tamed by the non-unitary dynamics, resulting in a decrease of the distance to a value which is independent of the initial state. That the distance does not vanish at late time can be attributed to the fact that the SC evolution and the quantum one do not evolve toward the same stationary state, as already discussed.

\begin{widetext}
	
	\begin{figure}[H]
		\centering
		\includegraphics[width=0.49\textwidth]{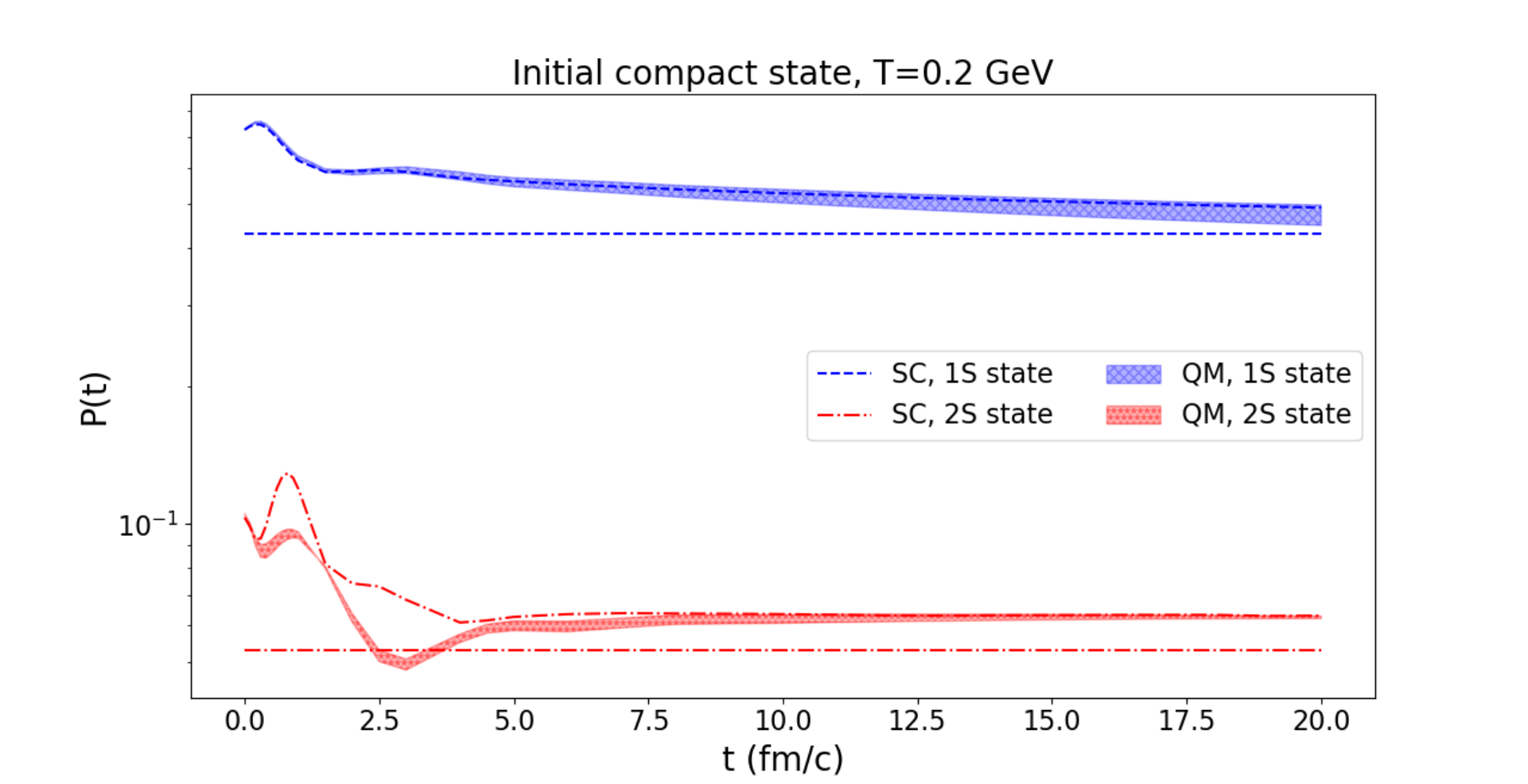}
		\includegraphics[width=0.49\textwidth]{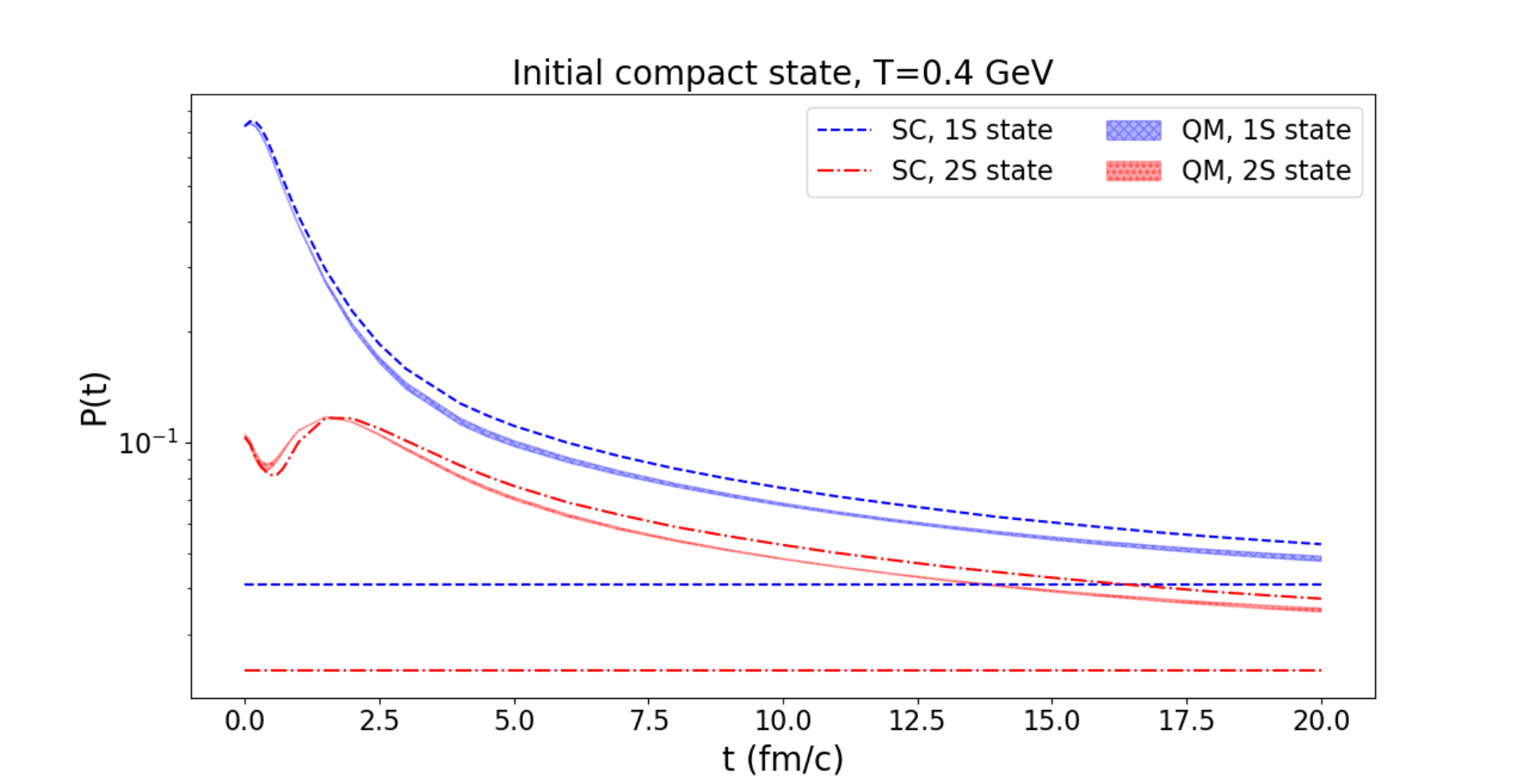} 
		\caption{QM vs. SC survival probabilities of 1S and 2S states for an initial compact state.  Left: for $T=0.2$ GeV. Right: $T=0.4$ GeV.}
		\label{fig:generation probabilties-compact state}
	\end{figure}
	
\end{widetext}

The survival probabilities for the 1S and 2S states are shown in Fig.~\ref{fig:generation probabilties-compact state}. Note that both the 1S and 2S states are ``present'' in the initial compact state. This initial superposition is responsible for the early-time Rabi-like oscillations that are clearly visible for $T=0.2$ GeV. The amplitude of these oscillations is larger for the 2S than for the 1S, in agreement with what was inferred about the early time dynamics from  the distances plotted in Fig.~\ref{fig:trace-distance-compact-state}. While they are well captured by the SC approximation for the 1S state, this is not so for the 2S at low temperature. At late time, one observes the same slight deviation between the QM and the SC description that we have already observed in the previous subsection for the evolution of the 1S state. 

Overall, the semiclassical approximation provides an excellent account of the evolution of the compact state, except at early time and low temperature where the unitary dynamics dominate and the associated quantum mechanical effects produce visible (but small) deviations between the two descriptions. 

\section{Discussion and conclusions}
\label{conclusions}

In this paper, we have conducted a comprehensive study, within a simple one-dimensional setting and an abelian plasma, with the objective of enhancing our comprehension of the conditions under which semiclassical approximations  can be applied, as well as the associated quantitative errors, in the context of charmonia production in AA collisions. This was achieved  by performing a comparative study between the exact quantum results obtained from the resolution of a one-dimensional Lindblad equation (\ref{Lindblad-map}) established in the quantum Brownian regime and the approximated semiclassical results obtained  by solving its associated  Fokker-Planck equation (\ref{Fokker}). In order to account for the various factors that can influence the validity of the semiclassical approximation, the equations were solved considering different medium temperatures and different (initial) states.

In general, we have found  that the semiclassical description reproduces the exact quantum results with remarkable accuracy, which lends credibility to the semiclassical approximation. To achieve this success, it is necessary to ensure that the constraints governing the derivation of the semiclassical equation (\ref{Fokker}), in particular the potential differentiability, are respected. In fact, there are several factors that can enhance such success of the semiclassical approximation. Our study has illuminated the crucial role of the non-unitary operators in classicalizing the initial quantum coherence and taming the deviations between the quantum and semiclassical descriptions. In other words, the inaccuracies in the semiclassical approximation of the unitary dynamics become inconsequential as soon as the non-unitary dynamics gradually takes over. For the various implementations considered in this work, this was shown to happen around $t\approx 1-2\,{\rm fm}/c$ for most observables. 

As the medium temperature increases, this non-unitary component of the dynamics becomes predominant at an earlier stage. Consequently, a more rapid convergence between the quantum and semiclassical descriptions is expected and indeed observed for some observables like the $\delta$ indicator considered in section \ref{sec_Decoherence and Classicalization} or the distance $d$ between the quantum and the semiclassical Wigner densities in the case of an initial compact state displayed in section \ref{sec:compact}. This can also be understood from the perspective of quantum decoherence which becomes more efficient at higher temperatures. However, the solution of the Lindblad-QME does not evolve toward the proper Gibbs-Boltzmann distribution as an asymptotic solution, an acknowledged feature identified in previous works which becomes more relevant at late time, especially at high temperatures. This leads to a slight systematic disagreement between the QM and the SC descriptions, and renders the temperature dependence less clear-cut to analyze for other observables such as the distance for an initial 1S state.

The agreement achieved for the quarkonia yield is excellent, with typical systematic errors of 10\% only. Deviations are observed to be relevant only at the early stages of the dynamics, particularly at low temperatures. In essence, the most challenging scenario for the semiclassical approximation pertains to the early-time behavior of the 2S probability, particularly that which is generated from a compact state within a QGP at low temperature, as illustrated on Fig. \ref{fig:generation probabilties-compact state}. However, for practical purposes, a low-temperature  QGP either results from a long-time evolution of an initially hot QGP (leaving plenty of time for the decoherence to act) or is produced in small and dilute systems, in which at most a single $c\bar{c}$ pair is produced, which can be efficiently described with a full QME. Consequently, it is not expected  that the less-favorable  scenario (low $T$, early stage) will have any significant impact on practical considerations. As for the late-time evolution, in view of the limitations of the present Lindblad - QME to reproduce the exact Gibbs-Boltzmann distribution, it may even argued --- in a slightly provocative way --- that the semiclassical description is superior to the quantum one.

In fact, the thermalization of an open quantum system as captured by the Lindblad equation, and the associated thermal distribution, remains a subject of active research. It is argued that the steady state in this case is the so called ``mean-force Gibbs state" which is expected to reduce to the Gibbs-Boltzmann distribution  only in the limit of an ultra-weak coupling system-medium, see e.g. \cite{Trushechkin:2021drp, Cresser:2021lop, Timofeev:2022tbl,tupkary2022fundamental}.

In a subsequent study, we intend to explore the asymptotic behavior of quarkonia states in greater depth within the framework of open quantum systems. In addition, in light of the restriction of the present study to the abelian case, a subsequent study is planned to be conducted in the non-abelian case. The objective of this subsequent study is to explore the potential role of color interactions in enhancing or diminishing the success of the semiclassical approximation, as observed in the abelian case.
For the sake of completeness, a similar study should also be performed in the quantum optical regime, which was not addressed in this work.

\begin{acknowledgments}
A.D.-H. acknowledges support from the Centre national de la recherche scientifique (CNRS/IN2P3) and the Région Pays de la Loire. The work of S.D. was supported by National Science Centre Poland under the Sonata Bis grant No 2019/34/E/ST2/00186. P.B.G. has received funding from the European Union’s Horizon 2020 research and innovation program under grant agreement STRONG – 2020 - No 824093.
\end{acknowledgments}

\appendix

\section{Superoperators}
In this appendix, we provide the explicit expressions for the $\mathcal{L}_{i}$ superoperators  in the abelian case: 
\begin{equation}
	\mathcal{L}_0=\frac{i}{M}\left(\partial_s^2-\partial_{s'}^2 \right),
\end{equation}
\begin{equation}
	\mathcal{L}_1=-iC_F\left(V(s)-V(s') \right),
\end{equation}
\begin{eqnarray}
	\mathcal{L}_2\mathcal{D}(s,s')&=&
	C_F \left[2\tilde{W}(0)-\tilde{W}(s) -\tilde{W}(s') +2 \tilde{W}\left(\frac{s+s'}{2}\right) \right.
	\nonumber\\ && \left. -2\tilde{W}\left(\frac{s-s'}{2}\right) \right]\mathcal{D}(s,s'),
\end{eqnarray}
\begin{eqnarray}
	\mathcal{L}_3\mathcal{D}(s,s')&=&
	\frac{C_F}{2 MT} \left[-\tilde{W}'(s)\partial_s-\tilde{W}'(s') \partial_{s'} -\tilde{W}'\left(\frac{s-s'}{2}\right) \right.
	\nonumber\\ && \left.
	(\partial_s-\partial_{s'})+\tilde{W}'\left(\frac{s+s'}{2}\right)(\partial_s+\partial_{s'})\right]\mathcal{D}(s,s'),\nonumber\\&&
\end{eqnarray}
and
\begin{eqnarray}
	\mathcal{L}_4\mathcal{D}(s,s')&=&
	\frac{C_F}{16 M^2T^2} \left[\tilde{W}''(0)(\partial_s^2+\partial_{s'}^2) + \tilde{W}''(s)\partial_s^2
	\right.
	\nonumber\\ && 	+\tilde{W}''(s')\partial_{s'}^2) 
	+ \tilde{W}'''(s)\partial_s + \tilde{W}'''(s')\partial_{s'} 
	\nonumber\\
	&&\left.+
	2 \left(\tilde{W}''(\frac{s-s'}{2})+\tilde{W}''(\frac{s+s'}{2})\right)\partial_s \partial_{s'}
	\right],\nonumber\\
	&&
\end{eqnarray}
where $\mathcal{D}(s,s')$  stands for the matrix element $\langle s| \mathcal{D} |s'\rangle$ with $s$ (or $s'$)  the relative coordinates of the  quarks. Note that we keep the Casimir factor $C_F$ in order to get phenomenologically relevant numbers in our simulations. $M$ is the mass of the heavy quark and $T$ the QGP temperature. One can easily check that each of these terms is norm preserving. The expression of $\mathcal{L}_3\mathcal{D}$ appears to be slightly different from the one derived in \cite{blaizot2018quantum}. The origin of such difference is that a so-called ``minimal set" of operators was identified in  \cite{Delorme:2024rdo} in a process aiming at regularizing UV divergences present in the higher derivatives of the imaginary potential $W$ and leading to a regularized imaginary potential $\tilde{W}$. This allows to absorb some terms of the original derivation into the definition of $\tilde{W}$, resulting in some simplification of the expressions. 

One of the main features of the QME is the possibility of approximating it by a semiclassical equation as shown in \cite{blaizot2018quantum}. This approximation  is based on the observation that the density matrix becomes eventually nearly diagonal in coordinate space. As for the $\mathcal{L}_1$ potential term $i\left(V(s)-V(s')\right)$, it is approximated as 
\begin{equation}
	i\left(V(s)-V(s')\right)
	\to i V'(r) \times y, 
	\label{eq:SCL1}
\end{equation}
which is the 1rst term in the semiclassical Wigner-Moyal equation for closed quantum systems~\cite{wigner1932quantum}. In this expansion, $r=\frac{s+s'}{2}$ while $y=s-s'$. With these variables one has  
\begin{eqnarray}
	\mathcal{L}_2\mathcal{D}(r,y)&=&
	C_F \left[2\tilde{W}(0)-\tilde{W}(r+\frac{y}{2}) -\tilde{W}(r-\frac{y}{2}) +2 \tilde{W}\left(r\right)
	\right. 
	\nonumber  \\  &&  \left. -2\tilde{W}\left(\frac{y}{2}\right) \right]\mathcal{D}(r,y),
	\nonumber  \\
	&=&
	-\frac{C_F}{4} \left[(\tilde{W}''(r)+\tilde{W}''(0)) y^2+ \mathcal{O}(W^{(4)}y^4)\right]
	\\ && \times \mathcal{D}(r,y).\nonumber  \\
\end{eqnarray}
Uniformly neglecting the $\mathcal{O}(W^{(4)}y^4)$ term is legitimate whenever it is small on the support of $\mathcal{D}$; this typically amounts to $\langle y^2 \rangle_\mathcal{D} \lesssim \frac{W^{(2)}}{W^{(4)}}\approx \mu_D^{-2}$,
where $\mu_D^{-1}$ (the inverse Debye mass) is the typical length scale appearing in the definition of the imaginary potential), a condition that is satisfied in the QBM regime for both (reasonable) initial and asymptotic state.
Similarly, one has
\begin{eqnarray}
	\mathcal{L}_3\mathcal{D}
	&=&-\frac{C_F}{2 MT} \left[\left(\frac{\tilde{W}'(r^-)+\tilde{W}'(r^-)}{2}- \tilde{W}'(r)\right)\partial_r
	\right. 
	\nonumber  \\  &&  \left. 	
	+\left(2\tilde{W}'\left(\frac{y}{2}\right)+\tilde{W}'(r^+) -\tilde{W}'(r^-) \right)\partial_y\right]\mathcal{D}
	\nonumber\\
	&\approx&-\frac{C_F}{2 MT} \left[\left(\tilde{W}''\left(0\right)+\tilde{W}''(r)\right) y\partial_y+\tilde{W}'''(r)\frac{y^2}{4}\partial_r\right]\mathcal{D},
	\nonumber  \\
\end{eqnarray}
where $r^\pm=r\pm\frac{y}{2}$.

\section{The regularized potential}
\label{sec:regpot}

As shown by Escobedo and Blaizot~\cite{blaizot2018quantum} and recalled in Sect.~\ref{sect:background}, the small $y$ expansion of the Lindblad QME followed by a Wigner transform results in a Fokker Planck equation whose drag and diffusion transport coefficients are expressed as derivatives of the original imaginary potential. In our work, we adopt this original derivation applying it to the minimal set of operators present in the QME and neglecting the contributions stemming from the $\mathcal{L}_4$ term.

\begin{widetext}
	
	\begin{figure}[H]
		\centering
		\includegraphics[width=0.45\textwidth]{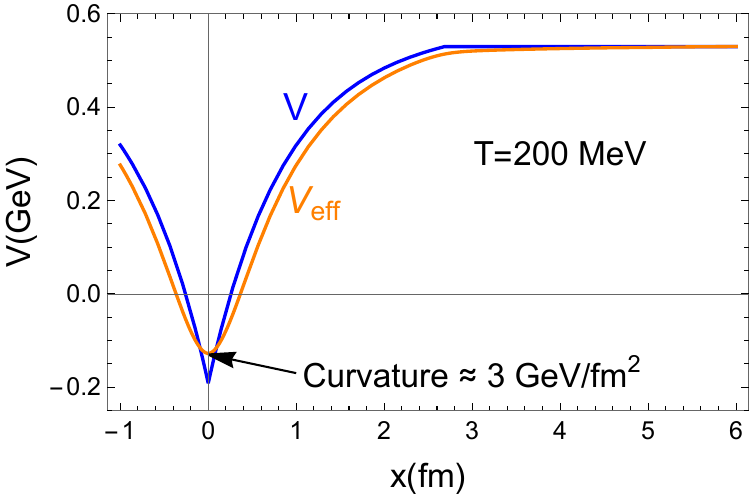}
		\hspace{1cm}
		\includegraphics[width=0.45\textwidth]{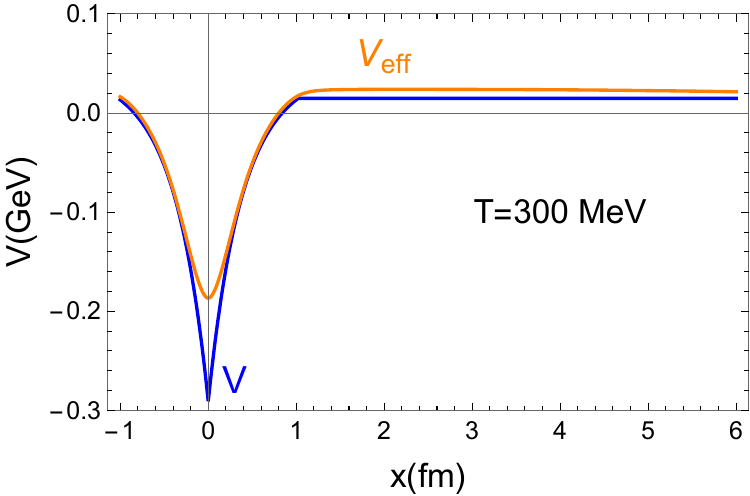}
		\caption{ A comparison between the native and regularized potentials for $T=0.2$ GeV and $T=0.3$ GeV.  }
		\label{fig:V-eff-illustration}
	\end{figure}
	
\end{widetext}

As for the unitary operators $\mathcal{L}_0$ and $\mathcal{L}_1$, the same procedure results in the standard Wigner - Moyal equation, truncated to its first order in $\hbar$ (i.e. retaining the contributions $\propto \hbar^0$). It should be noted that such expansion where $V(s)-V(s')$ is approximated by $V'(\frac{s+s'}{2})\times y$ requires that the real potential $V$ is differentiable. Since this is not the case at origin for the 1D potential derived in~\cite{Katz:2022fpb}, we have used instead, in the Langevin equation, the regularized potential $V_\mathrm{reg}$ defined following the method of Kelbg~\cite{AVFilinov:2003}:
\begin{equation}
	V_\mathrm{reg}(T;r):=-T \ln \rho_\mathrm{as}(T;r) +\mathrm{cst}(T),
\end{equation}
where $\rho_\mathrm{as}$ is the asymptotic quantum density evaluated by solving $\mathcal{L}\mathcal{D}_\mathrm{as}=0$. Adopting this regularization ensures that the density resulting from the FP evolution converges toward the equivalent quantum quantity as $\lim_{t\to +\infty} \rho_\mathrm{FP}\propto \exp(-V/T)$. At intermediate times, this prescription also reduces the discrepancy between the quantum and the SC treatments, as we have checked explicitly. In Fig.~\ref{fig:V-eff-illustration}, the regularized potential is illustrated and compared to the native 1D potential. One sees that they essentially differ close to the singular point of the native potential at origin. 

\section{Finer analysis of the distance}
\label{sec:ddrdp}

In this appendix, we complement the  analysis of the distance presented in Fig.~\ref{fig:distance} by discussing two particular features. First, the flattening of the distance toward constant values at late times seems to indicate that the system is approaching equilibrium, with however different asymptotic states associated to both quantum and semiclassical descriptions. Second, a clear hierarchy between the temperatures is lacking, although one would expect on general grounds \cite{joos2013decoherence,schlosshauer2007decoherence}   that decoherence should proceed faster and more efficiently at high temperature, as observed in Fig.~\ref{fig:Wigner-negativity}. 

\begin{widetext}
	
	\begin{figure}
		\centering
		\includegraphics[width=0.49\textwidth]{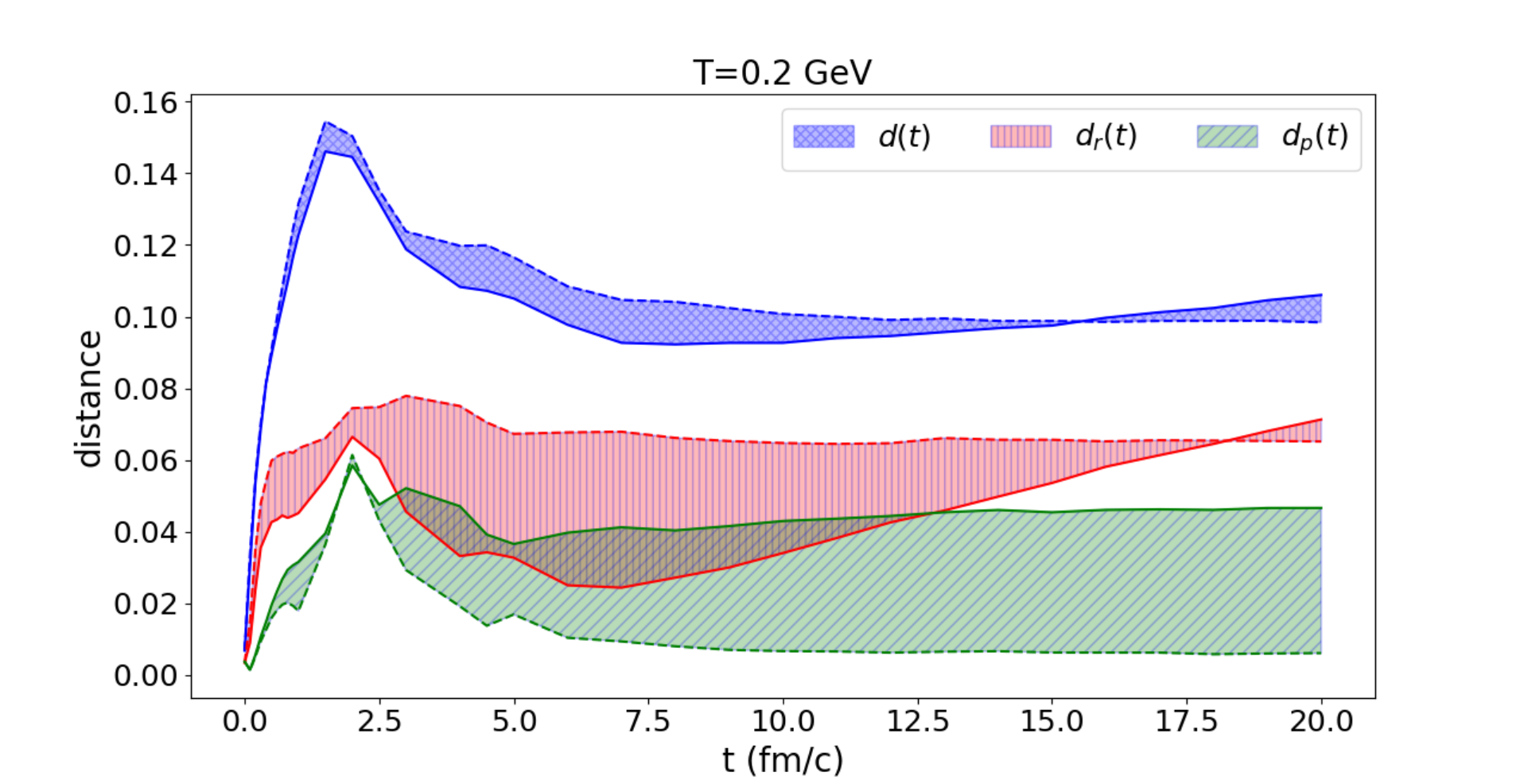}
		\includegraphics[width=0.49\textwidth]{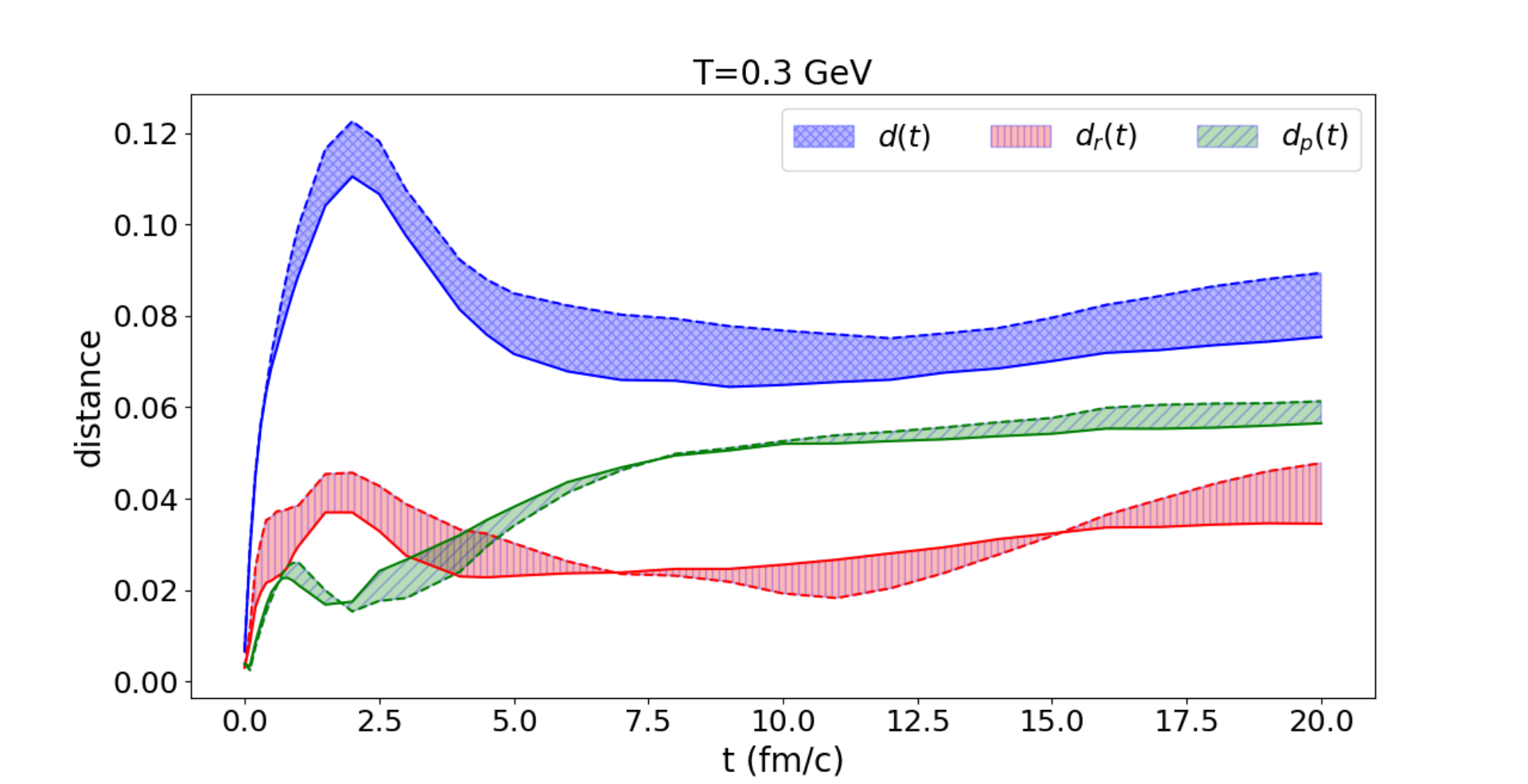}\par 
		
		\includegraphics[width=0.49\textwidth]{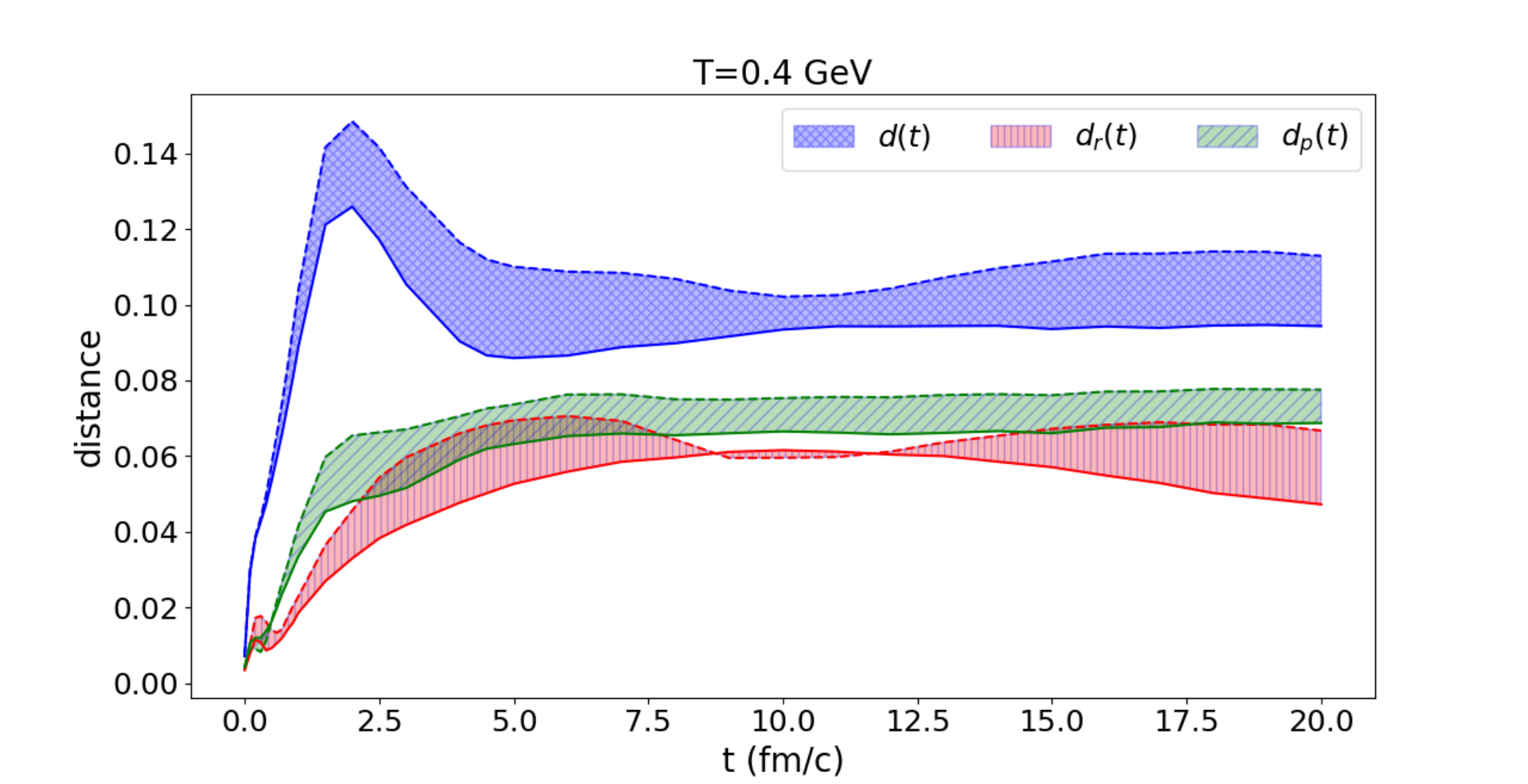}
		\includegraphics[width=0.49\textwidth]{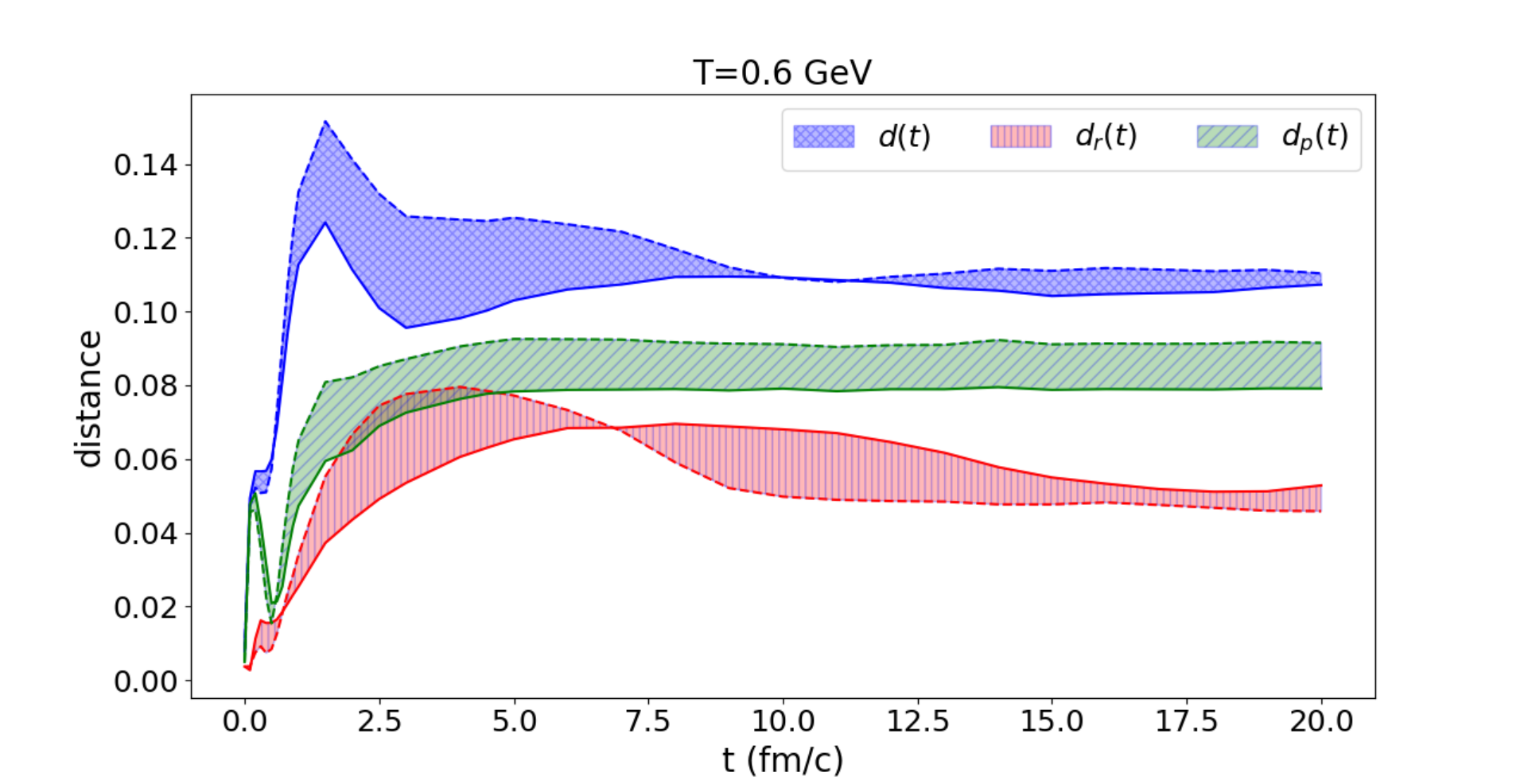}\par 
		\caption{Time evolution of the distance  $d(t)$  and its position and momentum contributions for  various QGP temperatures. The solid and dashed lines framing the bands respectively correspond to the QME evolution with and without $\mathcal{L}_4$ term.}
		\label{fig:trace-distance-L1-decomposed}
	\end{figure}
	
\end{widetext}

In order to investigate these features,  we   use the marginal distances
\begin{eqnarray}
	d_r\left(t\right)&=& \int\text{d}r\left|\rho_\mathrm{QM}\left(t,r\right)-\rho_\mathrm{SC}\left(t,r\right)\right|
	\nonumber\\ 
	d_p\left(t\right)&=& \int\text{d}p\left|\rho_\mathrm{QM}\left(t,p\right)-\rho_\mathrm{SC}\left(t,p\right)\right|, \label{eq:distance-r-p-decomposition}  
\end{eqnarray}
with $\rho\left(r\right)$ and $\rho\left(p\right)$ being respectively the position and momentum marginal distributions associated to the Wigner function $\mathcal{D}\left(r,p\right)$.  The position and momentum  distances  satisfy the inequalities $d_r(t)\leq d(t)$ and $d_p(t)\leq d(t)$,  respectively. These distances are plotted in  Fig.~\ref{fig:trace-distance-L1-decomposed} 
for various temperatures.

We notice distinct behaviors of these components. While $d_r$ is roughly independent of $T$, $d_p$ clearly grows with the temperature. Indeed $d_r>d_p$ throughout the evolution process at $T=0.2$ GeV. However, as temperature is increased to $T=0.3$ GeV, the dominance of $d_r$ is observed only at early times, with $d_p$ taking over rapidly. Finally, for the highest temperatures, $T=[0.4,0.6]$ GeV, $d_p$ dominates the distance from the outset. 

The late-time apparent stability observed for $d_r$ up to $20\,{\rm fm}/c$ is in fact a direct consequence of the large times scales associated with the spatial relaxation, as already seen for $\langle r^2 \rangle$ in Fig~\ref{fig:sqrt-r-squared}. For $t\gg 20\,{\rm fm}/c$, $\rho_{\rm SC}(r)$ is however assured to converge toward $\rho_{\rm QM}(r)$ due to the choice of the regulated potential $V_{\rm reg}$ adopted for the SC evolution and $d_r$ should decrease toward 0.  For a finite $t$, the position distance $d_r$ can be traced back to the higher quantum corrections which were neglected in the semiclassical equation. Given that both components of dynamics contain higher-order quantum corrections, it can be concluded that the position distance is not significantly affected by changes in medium temperature. Indeed, at low temperature, the primary source of discrepancies is the quantum corrections associated to the real potential $V(r)$ (see as well Fig.~\ref{fig:abs errors-new}), while at high temperature,  where $V(r)$ becomes flatter, the primary source of discrepancies is the quantum corrections stemming from the imaginary potential, $W(r)$.\footnote{It is worth to note that at $T=0.6$ GeV, the real potential is flat, so the unitary component of dynamics  is identical in the quantum and semiclassical descriptions. Consequently, the non-zero $d_r$ distance at this temperature, quantifies the discrepancies induced by the higher order quantum corrections in the non-unitary component of the dynamics. } Therefore, the decrease in one type of correction is compensated for by an increase in the other. This elucidates the absence of a manifest hierarchy between the different temperatures in the distance $d_r$ but also in the distances $d$ shown in Fig.~\ref{fig:distance}. 

As for $d_p$, we have already argued (see Fig.~\ref{fig:sqrt-p-squared}) that the quantum and semiclassical equations do not possess the same steady-state solution since the Lindblad equation (\ref{Lindblad-map}) does not relax to a Gibbs-Boltzmann distribution in the same manner as the Fokker-Planck equation. At late times, the distance $d_p$ is thus a  measure of the distance between these different steady states and remains non zero. It even increases as a function of temperature as observed in both Figs.~\ref{fig:sqrt-p-squared} and ~\ref{fig:trace-distance-L1-decomposed}. This is also in agreement with the increase of $\|\Delta\mathcal{L}_2\|$ observed on Fig.~\ref{fig:abs errors-new}.

\bibliographystyle{apsrev4-2}
\bibliography{QMvsSC}

\end{document}